\documentclass[twocolumn]{aastex63}
\received{\today}
\revised{}
\accepted{}
\usepackage{xcolor}
\usepackage{lineno}

\definecolor{mygreen}{rgb}{0.19,0.55,0.11}
\shorttitle{Simulations of Black Hole--Neutron Star Progenitors}
\shortauthors{Jiang~et~al.}
\begin{document}
\title{Simulations of the progenitors of black hole--neutron star gravitational wave sources}
\correspondingauthor{Wen-Cong Chen}
\email{chenwc@pku.edu.cn}
\author[0000-0002-2479-1295]{Long Jiang}
\affil{School of Science, Qingdao University of Technology, Qingdao 266525, China}
\affil{School of Physics and Electrical Information, Shangqiu Normal University, Shangqiu 476000, China}
\author[0000-0002-0785-5349]{Wen-Cong Chen}
\affil{School of Science, Qingdao University of Technology, Qingdao 266525, China}
\affil{School of Physics and Electrical Information, Shangqiu Normal University, Shangqiu 476000, China}
\author[0000-0002-3865-7265]{Thomas M. Tauris}
\affiliation{Department of Materials and Production, Aalborg University, Skjernvej 4A, DK-9220~Aalborg {\O}st, Denmark}
\author[0000-0002-4470-1277]{Bernhard M\"{u}ller}
\affiliation{Monash Centre for Astrophysics, School of Physics and Astronomy, 11 College Walk, Monash University, Clayton VIC 3800, Australia}
\author[0000-0002-0584-8145]{Xiang-Dong Li}
\affil{Department of Astronomy, Nanjing University, Nanjing 210023, People's Republic of China}
\affil{Key Laboratory of Modern Astronomy and Astrophysics, Nanjing University, Ministry of Education, Nanjing 210023, People's Republic of China}
\begin{abstract}
Recent discoveries of gravitational wave (GW) events most likely originating from black hole (BH) + neutron star (NS) mergers reveal the existence of BH+NS binaries. The formation of BH+NS binaries and their merger rates through isolated binary evolution have been investigated extensively with population synthesis simulations. A detailed stellar evolution modelings of the formation of this population, however, is missing in the literature. 
In this work, we perform the first complete 1D~model of  more than 30 BH+NS progenitor systems which are calculated self-consistently until the iron core collapse with infall velocity exceeds 1000 km~s$^{-1}$. 
Focusing on  the progenitors of BH-NS GW sources, we apply the \texttt{MESA} code starting from a post-common envelope binary with short  orbital period ($<1\;{\rm day}$) consisting of a BH and a zero-age main-sequence helium star that experiences stable mass transfer.
The (ultra-)stripped supernova explosion is subsequently modelled using a semi-analytic method to reveal final remnant masses and momentum kicks.
Three example systems (A, B, and C) eventually evolve into BH+NS binaries with 
component masses of $(M_{\rm BH},\,M_{\rm NS}) =(8.80,\,1.53)$, $(8.92,\,1.45)$, and $(5.71,\,1.34)\;M_\odot$, respectively.
These NS masses could be significantly larger depending on the exact mass cut during the supernova explosion.  
These BH+NS systems are likely to merge and produce GW events within a Hubble time. System~C is a potential progenitor of a GW200115-like event, while Systems~A and B are possible candidates for a GW200105-like event and may represent the final destiny of the X-ray binary SS433.    
\end{abstract}
\section{Introduction}
Compact objects are fossils and probes for testing stellar and binary evolutionary theory.
Among double compact object systems, neutron stars (NSs) are generally observed as radio pulsars accompanied by either a white dwarf or another NS. Despite their existence predicted by population synthesis \citep[e.g.][]{kruc18,csh+21}, so far no black hole (BH)+NS binaries have been detected in our Galaxy. Current and future radio observations by the Five-hundred-meter Aperture Spherical radio Telescope and the Square Kilometre Array \citep{liu14,shao18,welt20}, however, may discover such a BH+NS binary which would open new possibilities for testing theories of gravity and BH physics \citep{liu14,welt20}. Another, and possibly more promising, channel is detection of a Galactic BH+NS binary using low-frequency gravitational wave (GW) detectors such as LISA \citep{LISA22}, TianQin \citep{lcd+16} or Taiji \citep{rgcz20}, that are able to track the continuous emitted GW signal for $\mathcal{O}( 10^5\;{\rm yr}$) prior to their merger event.

In 2020, the third observing run of the LIGO–Virgo-KAGRA detector network discovered the GW signals from two extragalactic BH+NS mergers: GW200105 and GW200115 \citep{abbo21}. The BH and NS component masses were inferred to be $8.9^{+1.2}_{-1.5}\;{\rm M}_{\odot}$ and $1.9^{+0.3}_{-0.2}\;{\rm M}_{\odot}$ (GW200105), and $5.7^{+1.8}_{-2.1}\;{\rm M}_{\odot}$ and $1.5^{+0.7}_{-0.3}\;{\rm M}_{\odot}$ (GW200115). Since then, a few other BH+NS merger candidates have been announced \citep{aaa+21}, including the marginal events GW190917 ($9.7^{+3.4}_{-3.9}\;{\rm M}_\odot$, $2.1^{+1.1}_{-0.4}\;{\rm M}_\odot$) and GW$190426$ ($5.7^{+2.0}_{-2.5}\;{\rm M}_\odot$, $1.5^{+0.8}_{-0.5}\;{\rm M}_\odot$) which resemble GW200105 and GW200115 in terms of component masses, respectively.

Whereas the empirically inferred merger-rate density in the local Universe of such BH+NS systems \citep[$\mathcal{R}_{\rm obs}\approx 45^{+75}_{-33}~\rm Gpc^{-3}\,yr^{-1}$, based on the first two BH+NS GW events,][]{abbo21} can be reproduced from population synthesis relatively easily due to the many poorly constrained input parameters, a detailed binary stellar evolution calculation of their progenitors is still missing. Here, in this work we present the first such calculation using the \texttt{MESA} code.

The main formation channel resulting in BH+NS binaries remains somewhat controversial. 
A number of different scenarios have been suggested \citep[see][for a review]{tv23}, however it is unclear which one is the main channel. 
The classic formation channel \citep[see e.g. Figure~15.38 in][]{tv23} includes both a stable mass-transfer stage and a common envelope (CE) phase \citep{heuv73,bhat91,voss03,taur06,belc08,mand22}. The original zero-age main sequence (ZAMS) binary consists of a massive star (the primary) and a somewhat less massive star (the secondary) in an orbit with a separation of $0.15-50{\rm ~AU}$.

After the primary has lost all of its hydrogen-rich envelope via Case~B mass transfer (Roche-lobe overflow, RLO), it experiences a core-collapse supernova (SN) and forms a BH. (Alternatively, a NS may form first in case of inverse formation order of the BH and the NS.) If the binary can keep bound during the SN, the secondary star subsequently also fills its RL and initiates mass transfer onto the BH. However, often this phase of mass transfer is dynamically unstable,
mainly due to a large mass ratio between the two stellar components, in which case the binary system enters a CE phase \citep{ijc+13} --- see, however, e.g. \citet{vpd17} for the possibility of producing BH+BH or BH+NS merger progenitors via stable RLO.  If the CE is successfully ejected, a close binary remains consisting of a BH and a naked helium core (Wolf-Rayet star), which might be observed as a system like Cyg~X-3 \citep{vcg+92,belc13}. Subsequently, the helium star (He~star) evolves and expands during helium shell burning, thereby triggering a stable phase of so-called Case~BB mass transfer onto the BH \citep{dpsv02,taur15}, until it collapses and leaves behind a NS in an ultra-stripped SN explosion \citep{taur13,taur15}. 

The formation of BH+NS binaries and their merger rates through isolated binary evolution have been investigated extensively with population synthesis simulations \citep[e.g.][]{tutu93,frye99,voss03,domi15,giac18,kruc18,neij19,belc20,shao21,broe21}.
For a review on merger rates from different simulations, see e.g. \citet{man22}.
For this method to work, fast calculations are needed which therefore make use of simple fitting formulae to mimic stellar evolution and binary interactions. In contrast, we present here the first detailed stellar evolution model of a BH+NS progenitor system based on self-consistent calculations of BH+He~star systems until the core collapse of the He~star. Our work can be considered a continuation of our similar calculations for double NS system progenitors \citep{jian21}.

\begin{table*}
\begin{center}
\centering
\caption{Initial and final parameters of three BH+He~star binaries\label{tbl-1}.}
\begin{tabular}{ccccccccccccc}
\hline\hline\noalign{\smallskip}
System & $M_{\rm BH,i}$ & $M_{\rm He,i}$ & $P_{\rm i}$ & $M_{\rm BH,f}$ & $M_{\rm He,f}$ & $P_{\rm f}$ & $M_{\rm He,f}^{\rm env}$ & $M_{\rm CO}$ & $M_{\rm NS}^\dagger$ & $E_{\rm SN}$ & $v_{\rm kick}$ & $P_{\rm merge}$\\
 & ${\rm M}_\odot$ & ${\rm M}_\odot$ & ${\rm day}$   & ${\rm M}_\odot$& ${\rm M}_\odot$ & ${\rm day}$ & ${\rm M}_\odot$ & ${\rm M}_\odot$ & ${\rm M}_\odot$  & $10^{51}{\rm erg}$  & ${\rm km~s}^{-1}$ & \%\\
\hline\noalign{\smallskip}
A  & 8.80 & 6.00 & 0.20 & 8.80$^\ddagger$ &  4.55  & 0.25 & 1.28 & 3.27 & 1.53  & 1.5 & 790 & 38.9\\
B  & 8.80 & 4.70 & 0.08 & 8.92 &  2.78  & 0.18 & 0.55 & 2.23 & 1.45  & 1.0 & 465 & 71.6\\
C  & 5.70 & 3.30 & 0.20 & 5.71 &  1.91  & 0.56 & 0.26 & 1.65 & 1.34  & 0.6 & 260 & 58.4\\
\hline\noalign{\smallskip}
\end{tabular}
\end{center}
\centering
\rm{\footnotesize{Notes: $\dagger$: The NS masses listed here are lower limits which result in slightly longer merger timescales and lower merger probabilities.\\
$\ddagger$: During the evolution, the BH accrected $\sim0.001~{\rm M}_\odot$ from the donor star, i.e., $M_{\rm BH,f}-M_{\rm BH,i}\simeq0.001~{\rm M}_\odot$.}}
\end{table*}

\section{Stellar evolution code}
To investigate the evolutionary history of NS+BH progenitors of GW200105/GW200115-like events, we simulated more than 30 BH+He~star binaries using the \texttt{MESA} code module \texttt{MESAbinary} \citep[version r$-$12778]{paxt11,paxt13,paxt15,paxt18,paxt19}. The parameters for three of them are presented in Table~\ref{tbl-1}. The evolution begins with a circular binary system with an initial orbital period of $P_{\text{i}}$, containing a BH with an initial mass of $M_{\text{BH,i}}$ and a helium ZAMS (He-ZAMS) donor star with an initial mass of $M_{\text{He,i}}$ and chemical compositions of $Y=0.98$ and $Z=0.02$. Most of the input physical  parameters used in the simulation are similar to those in our previous work on a double NS system progenitor \citep{jian21}.

Following \citet{lang91}, we adopt a mixing-length parameter, $\alpha=l/H_{\rm p}=1.5$, where $l$, and $H_{\rm p}$ denote the mixing length and the local pressure scale height, respectively. 
For the wind mass-loss rate of the He~star, we use the ``Dutch'' prescription \citep{gleb09} with a scaling factor of 1.0, and Type\,2 opacity were applied for $Z=0.02$.  We adopt $min_{-}mdot_{-}for_{-}implicit=10^{-10}\;{\rm M_\odot\,yr^{-1}}$. If the mass-transfer rate is lower than this limitation, the mass transfer is computed explicitly, 
avoiding many iterations when the He~star decouples from its Roche lobe (RL). 

We assume that the fast wind leaves the He star carrying its specific orbital angular momentum. We neglect wind accretion onto the BH because Bondi-Hoyle-like accretion in this case is inefficient \citep[see e.g.][and references therein, for discussions]{taur17}.
This is simply because the radius of gravitational capture of the BH: $r_{\rm acc}\sim 2GM_{\rm BH}/v_{\rm wind}^2$, where $v_{\rm wind}\sim 3000\;{\rm km\,s}^{-1}$ for a $6-8\;M_\odot$ He~star \citep{Vink17}, is significantly smaller than the orbital separation, $a$, which leads to a small accretion efficiency factor, $f_{\rm acc}\simeq \pi r_{\rm acc}^2/(4\pi a^2)$. As an example for our simulated System~A (see also Table 1), we find $r_{\rm acc}\simeq 0.37\;R_\odot$ and $a\simeq 3.5\;R_\odot$ , which leads to $f_{\rm acc}\simeq 0.0028$. Thus prior to Case~BB RLO, the BH would only accrete about $0.0035\;M_\odot$ of the $\sim 1.27\;M_\odot$ lost in a fast wind from its He-star companion.

The mass-transfer rate ($|\dot{M}_{\rm tr}|$) from the He~star is calculated using the ``kolb'' mass-transfer scheme based on optical thick RLO \citep{kolb90}.
We adopt an accretion efficiency $f=0.5$ and Eddington-limited accretion onto the BH, such that $\dot{M}_{\text{BH}}=\text{min}(-0.5\dot{M}_{\text{tr}}, \,\dot{M}_{\text{Edd}})$, and
use the prescription of \citet{pods03} for the Eddington accretion rate $\dot{M}_{\text{Edd}}$. 
The orbital evolution during RLO is solved using the ``isotropic re-emission'' model \citep{bhat91,tv23} which assumes that the excess material in unit time ($|\dot{M}_{\text{tr}}|-\dot{M}_{\text{BH}}$) 
is re-emitted as an isotropic fast wind with the specific orbital angular momentum of the BH. Orbital angular momentum loss due to GW radiation is included, although here its effect is negligible.

Different from \citet{jian21}, in the current work we use the nuclear network \texttt{mesa235.net} (which includes 235 nuclei) for the nuclear reactions. Additionally, element diffusion with the coefficients prescription of \citet{stan16} is applied when the time step exceeds 1~yr.

Given the challenging numerical modelling of the final stages of the evolution of the donor star just prior to core collapse (where the central density and temperature increase by several orders of magnitude),
the numerical modeling is divided into three stages (see also Table~\ref{tbl-2}), similar to \cite{jian21}:\footnote{See further details in \citet{jian21}. Our inlists are available at doi: \url{10.5281/zenodo.7549340}} 
\begin{itemize}
\item [(i)] Modelling from He-ZAMS to (almost) the end of C-burning and before the ignition of neon. 
In practice, this corresponds to a central temperature, $8.8\leq\log T_c<9.1$, 
and a central mass fraction of carbon, $f_{\rm C}\le 0.2\%$, 
at about $\sim 10-100$ years prior to core collapse.

\item [(ii)] Continuation until the central temperature reaches $\log T_{\rm c}=9.85$. At this point, the mass of the Si/Fe/neutron-rich core (the outer boundary of an element core is defined with a threshold mass fraction of $30\%$) approaches its maximum value, which occurs $\leq10^{-5}{\rm ~yr}$ (a few minutes) prior to core collapse.
\item [(iii)] Early core-collapse stage until the in-fall velocity of the iron core rises to $1000\;{\rm km\,s}^{-1}$, and the core collapse supernova (CCSN) takes place. 
\end{itemize}

\begin{table*}
\begin{center}
\centering
\caption{Key time epochs in the simulations of the BH+He~star systems.\label{tbl-2}}
\begin{tabular}{cccccc}
\hline\hline\noalign{\smallskip}
\footnotesize{System} & \footnotesize{Onset of Case~BA RLO} & \footnotesize{Onset of Case~BB RLO} & Restart~I& Restart~II & CCSN\\
 & $t_1$ & $t_2$  & $t_3$  & $t_4$ &$t_*$\\
 & ($t_*-t_1$)  & ($t_*-t_2$)  & ($t_*-t_3$)  & ($t_*-t_4$)  & \\
\noalign{\smallskip}  
\hline\noalign{\smallskip}
A  & ---          & 0.87~Myr      &  0.872310~Myr &  0.8723186909770~Myr &  0.8723186909819~Myr\\           
   & ---          & (2300~yr)     &  (8.3~yr)     &  (2.6~min)           & ---\\
B  & 0.14~Myr     & 1.11~Myr      & 1.132975~Myr  & 1.1330156652611~Myr  & 1.1330156652647~Myr\\         
   & (0.99~Myr)   & (23\,000~yr)   & (40.5~yr)     & (1.9~min)            & ---\\
C  & ---          & 1.67~Myr      & 1.704938~Myr  & 1.7050658509765~Myr  & 1.7050658509850~Myr\\
   & ---          & (35\,000~yr)   & (128.2~yr)    & (4.5~min)            & ---\\
\noalign{\smallskip}  
\hline\noalign{\smallskip}
\end{tabular}
\end{center}
\end{table*}

\section{Results}\label{sec:results}
In a binary system, the initial ZAMS stellar mass must be $M\ga 10-12\;M_{\odot}$ in order to form a NS \citep{tv23}, corresponding to an initial He-star mass of above $\sim 2.4-2.8\;M_{\odot}$ \citep{taur15,woo19}.
The upper ZAMS mass limit for leaving behind a NS is much more complex due to ``islands of explodability'' \citep{ujma12,sew+16,mhlc16,ecf+19}. Simply taking $25\;M_\odot$ as a typical upper mass limit, corresponding to a star producing an initial He-core mass of $\sim 9.0\;M_\odot$ \citep[e.g.][]{swh18}, we therefore simulate He-star masses in the range of 2.8 to $9.0\;M_{\odot}$  systematically, and a couple of systems with higher He-star masses of $10\;M_\odot$ and $15\;M_\odot$ for further discussions. In a future paper, we plan to explore He~stars with even higher masses that evolve to produce BH+BH systems.
For further discussions of the final outcome of evolved He~stars, see detailed work by \citet{woo19}.
To produce BH-NS systems that will merge within a Hubble time, we apply here a short initial orbital period of less than 1~day.

Employing the $\texttt{MESA}$ code, we successfully simulate the evolution of several BH+He~star binaries with different initial component masses and orbital periods until the infall velocity of the iron core reaches 1000 $\rm km\,s^{-1}$. In Table~\ref{tbl-1}, we list the initial parameters of three selected systems. The corresponding final binary parameters at the onset of the CCSN, {as well as our modelled post-SN parameters of explosion energy, NS mass and imparted SN kick,} are also given. It is clear that Systems~A and B are the progenitor candidates of GW200105-like events (see discussions in Sections~\ref{subsec:massive_He-stars} and \ref{subsec:merger-times}), while System~C might be a progenitor of the GW200115-like event, according to the inferred component masses of these two GW events \citep{abbo21}.

\begin{figure}
\includegraphics[trim=1cm 9cm 5cm 0.3cm, clip=true, width=0.74\textwidth, angle=0]{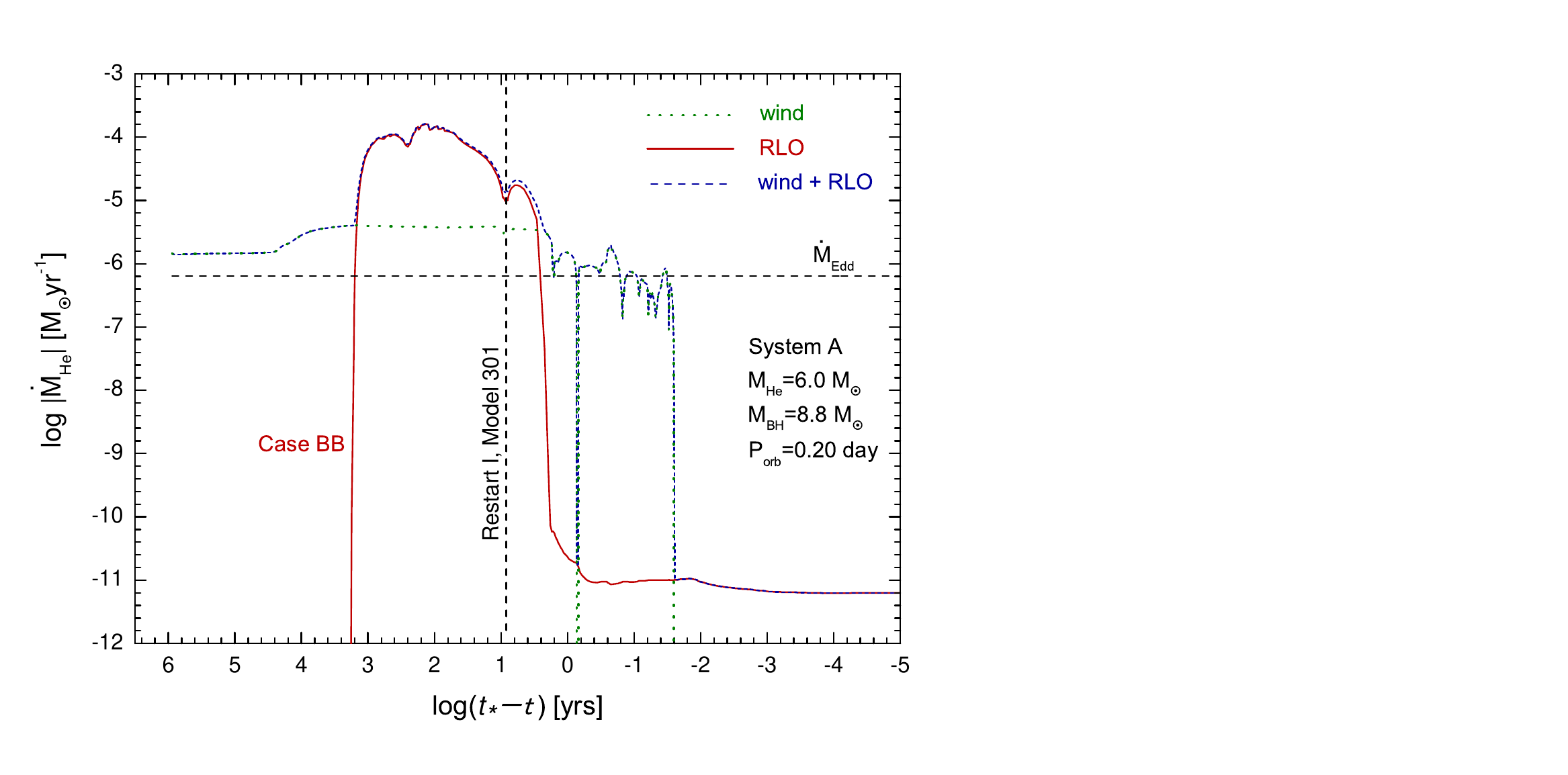}
\includegraphics[trim=1cm 9cm 5cm 0.9cm, clip=true, width=0.74\textwidth, angle=0]{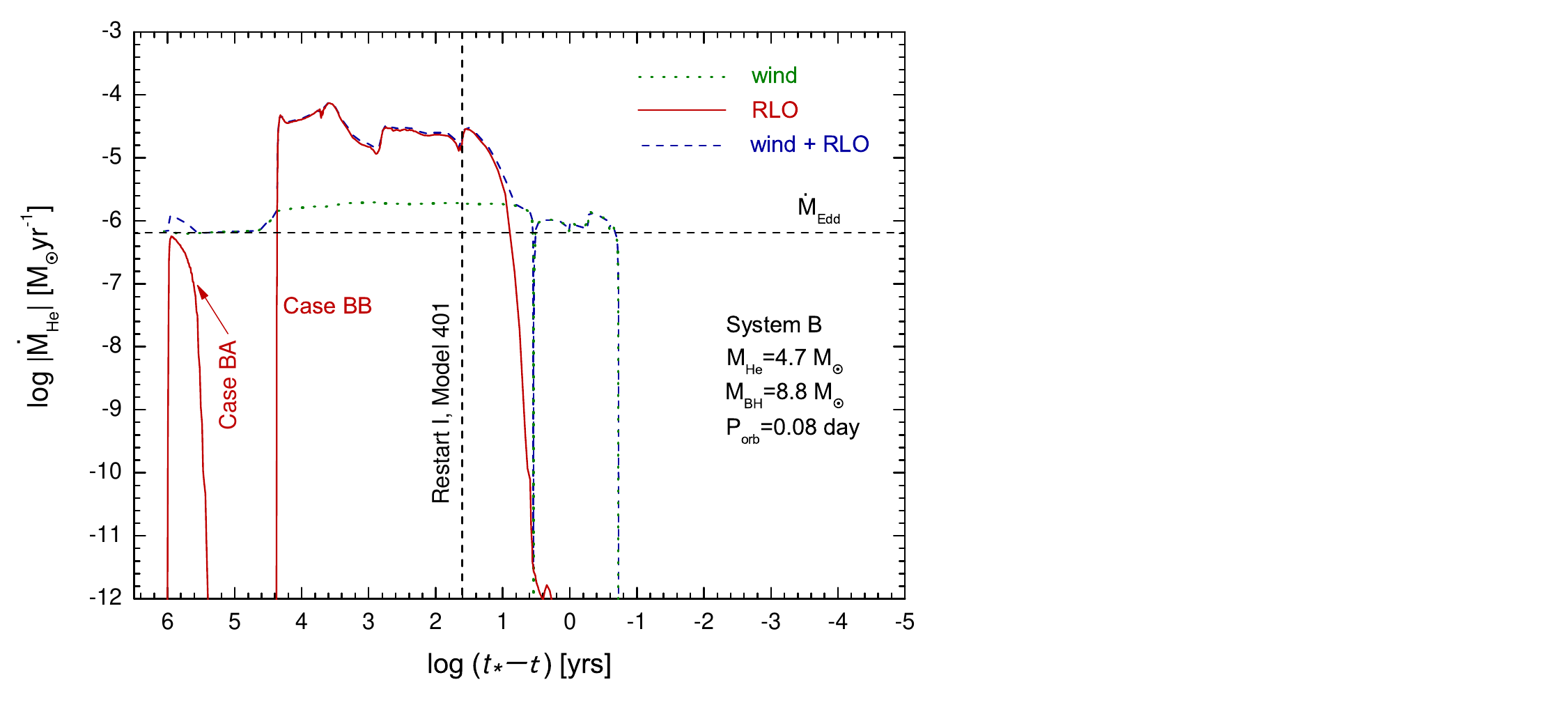}
\includegraphics[trim=1cm 9cm 5cm 0.9cm, clip=true, width=0.74\textwidth, angle=0]{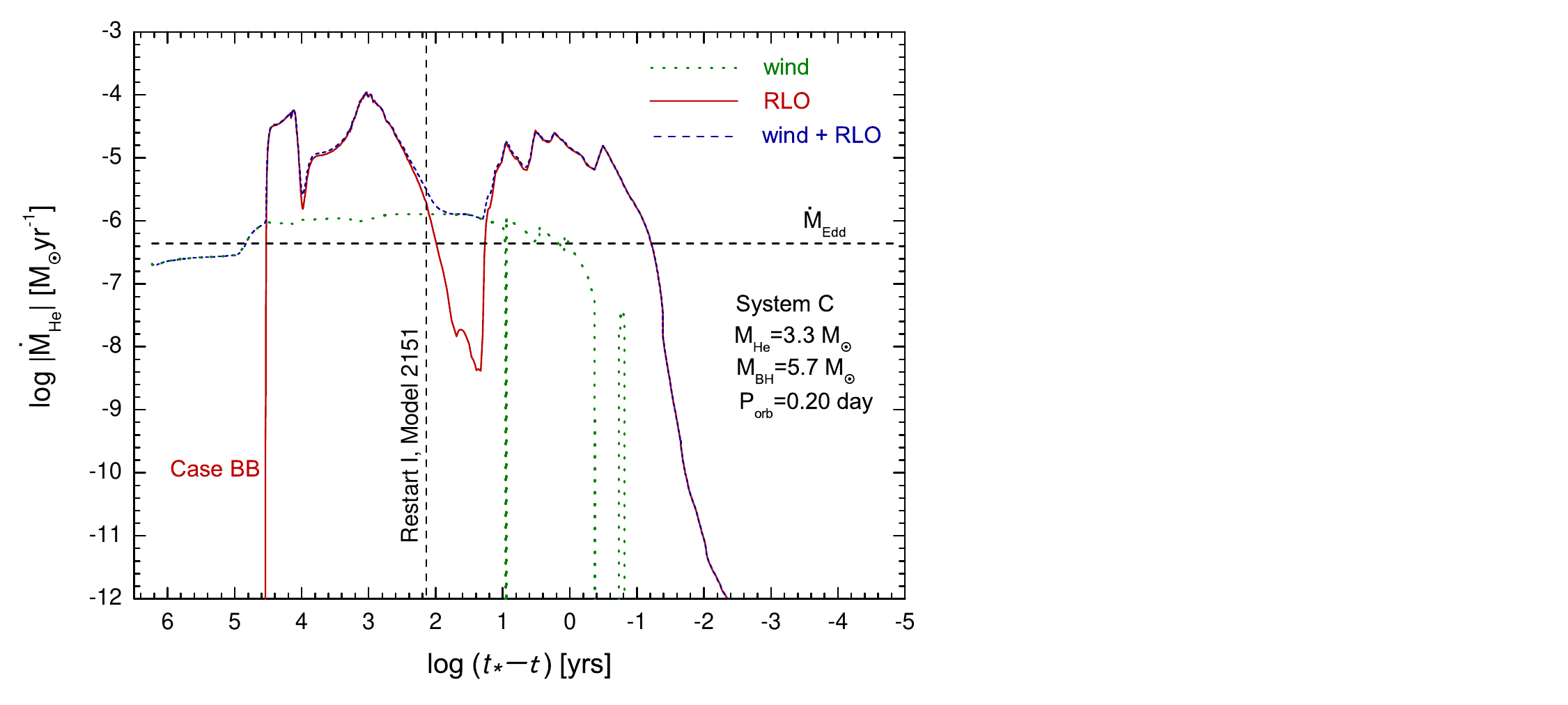}
\caption{Evolution of mass-loss rate of the He~star donors vs. remaining time until CCSN. The top, middle, and bottom panels illustrate System~A, B and C, respectively (same order for following figures). The red solid lines denote the effect of RLO while the blue short-dashed lines denote the total mass loss caused by wind and RLO combined. The horizontal and vertical black dashed lines display the Eddington accretion limit of the accreting BH and the transitions from stage~(i) to stage~(ii) (Restart~I), respectively. See text for a discussion of $|\dot{M}_{\rm He}|< 10^{-10}\;{\rm M}_\odot\,{\rm yr}^{-1}$.} 
\label{fig:mass-loss-rate}
\end{figure}

In Table~\ref{tbl-2}, we list some important times (stellar ages) of the simulations of the three systems as follows: the onset of Case~BA ($t_1$) ( i.e. mass transfer initiated during core helium burning) 
and Case~BB ($t_2$) RLO \citep[ i.e. after ignition of helium shell burning, ][]{hab86}, the transition times from stage~(i) to stage~(ii) ($t_3$, Restart~I) and from stage~(ii) to stage~(iii) ($t_4$, Restart~II). 

Figure~\ref{fig:mass-loss-rate} shows the evolution of the three He~star donors in the mass-loss rate  vs. remaining time ($t_*-t$) plot until core collapse, where $t_*$ is the total simulated stellar age from He-ZAMS to CCSN.
The mass-loss rate caused by RLO and wind are displayed by the red solid line and green dotted line, respectively, while the total mass loss caused by both wind and RLO is indicated by the the blue dashed line.
The Eddington accretion limit of the accreting BH is  illustrated as the black dashed line while the vertical dashed black line denotes Restar~I.

\begin{figure}
\centering
\vspace{0.7cm}
\includegraphics[trim=1cm 0cm 1cm 1.3cm, clip=true, width=0.43\textwidth, angle=0]{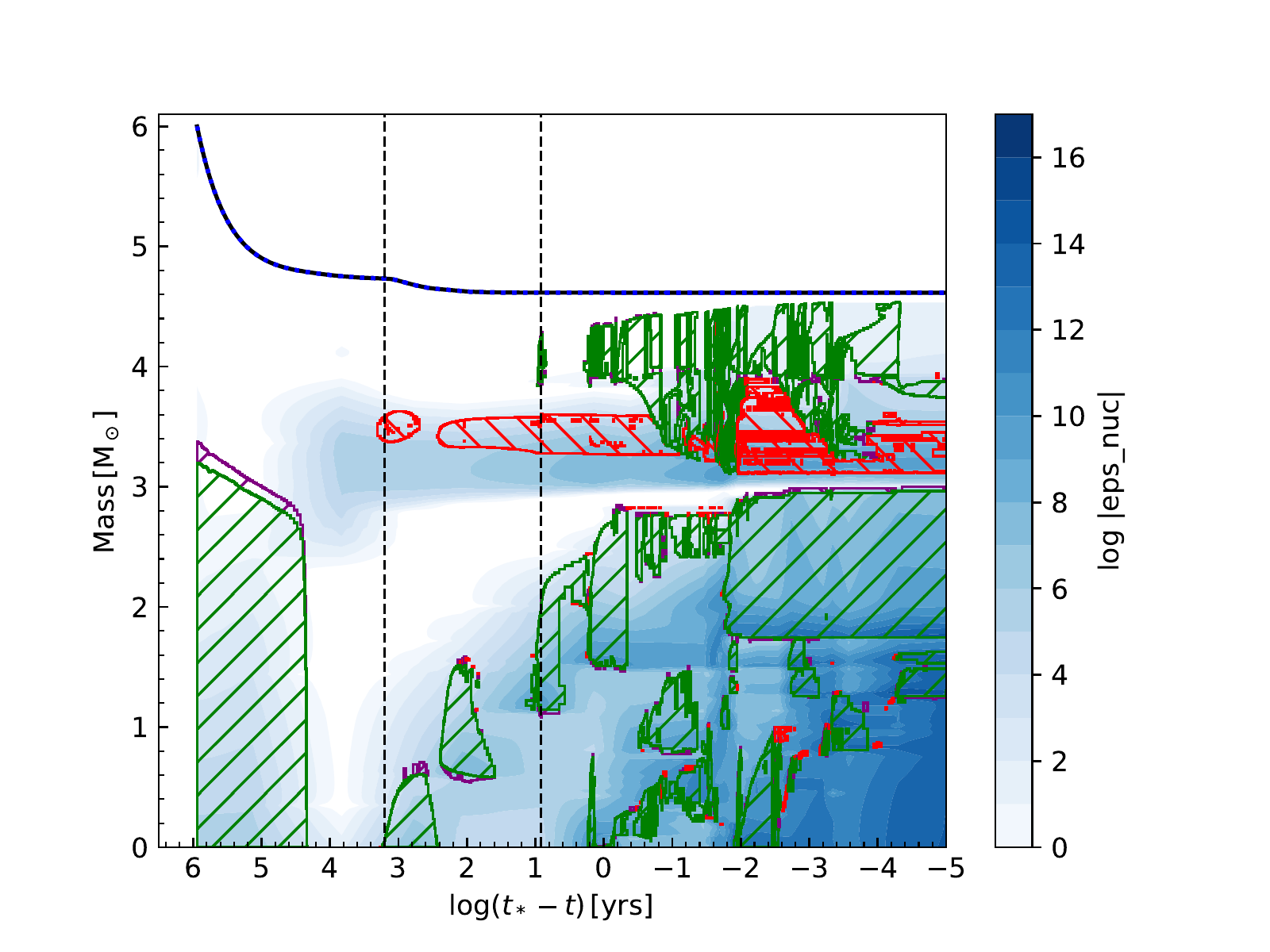}\vspace{-0.05cm}
\includegraphics[trim=1cm 0cm 1cm 1.3cm, clip=true, width=0.43\textwidth, angle=0]{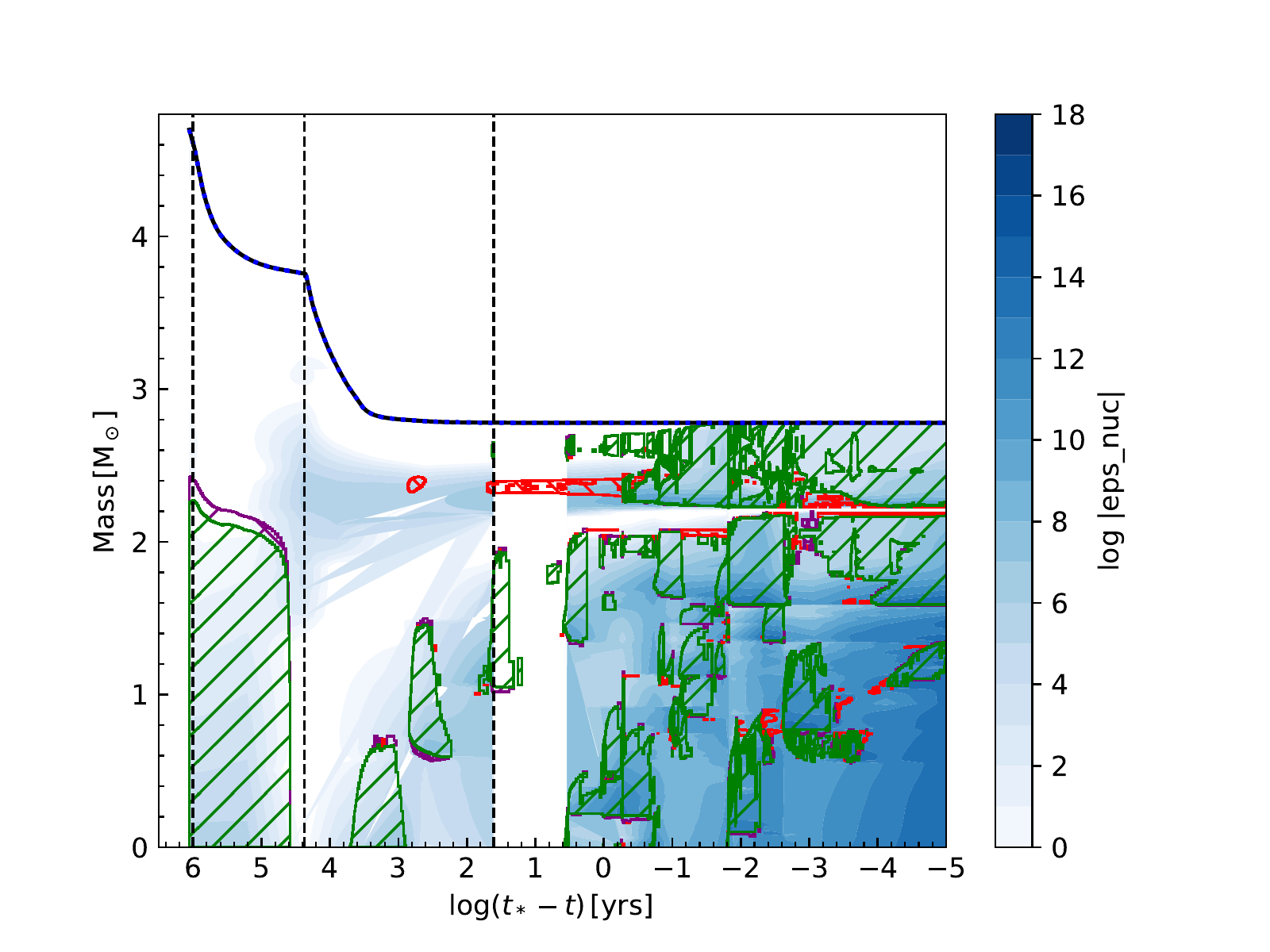}\vspace{-0.05cm}
\includegraphics[trim=0.8cm 0cm 1cm 1.3cm, clip=true, width=0.43\textwidth, angle=0]{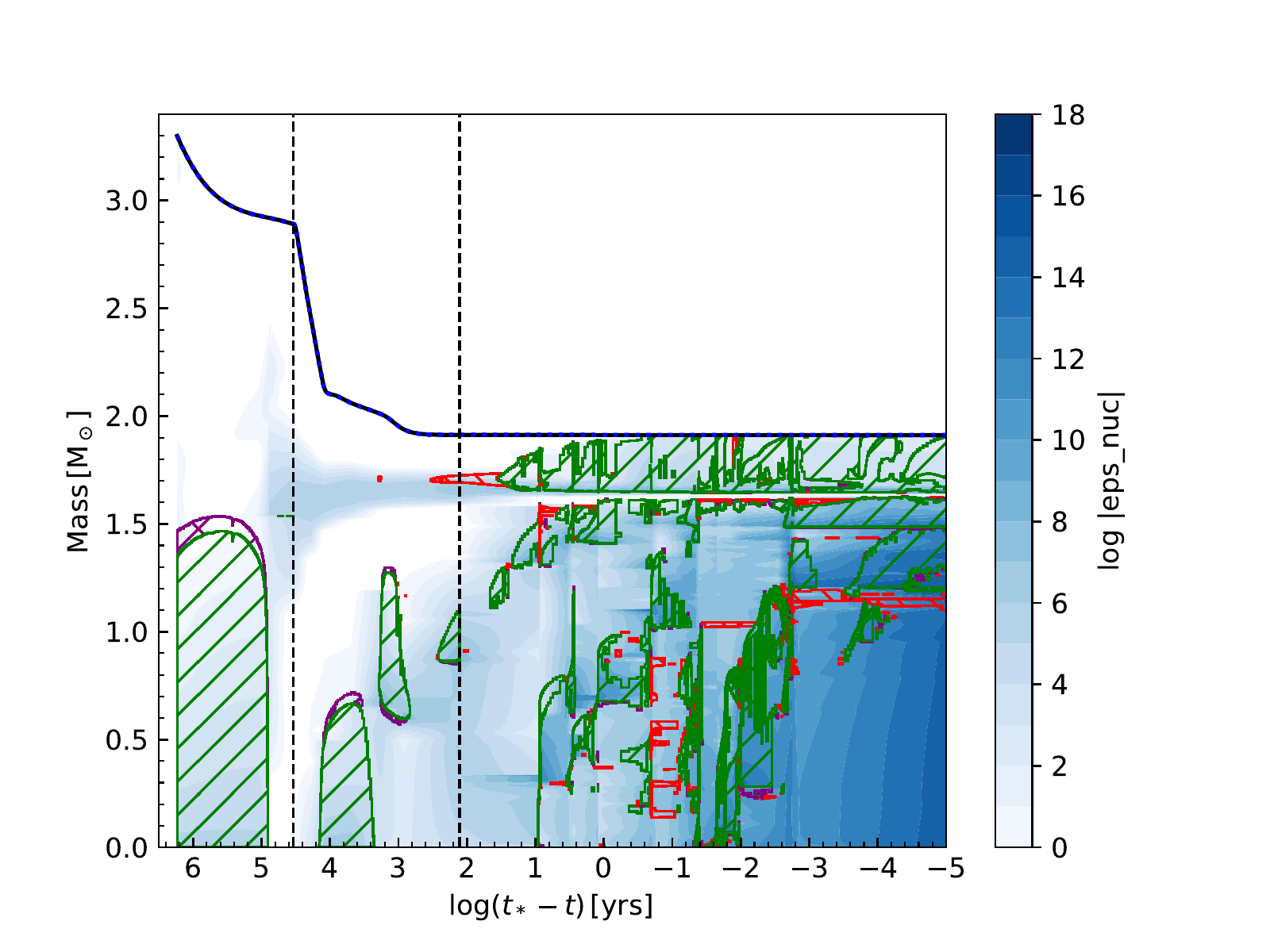}\vspace{-0.0cm}
\caption{Kippenhann diagram of the three donor stars (A, B and C). 
{The plot shows cross-sections of the He~stars in mass-coordinates, from the center to the surface, along the y-axis, as a function of remaining stellar age until core collapse on the x-axis.}
The nuclear energy production rate is shown as the intensity of the blue color (vertical scale at right). 
The  green color indicates convection while the red color hatched areas denote semi-convection. The vertical dashed lines mark, from left to right, the onsets of Case~BA RLO (in System~B only), Case~BB RLO, and Restart~I.
} 
\label{fig:kippenhann}
\end{figure}

\begin{figure}
\centering
\includegraphics[trim=1cm 9cm 3cm 0.6cm, clip=true, width=0.8\textwidth, angle=0]{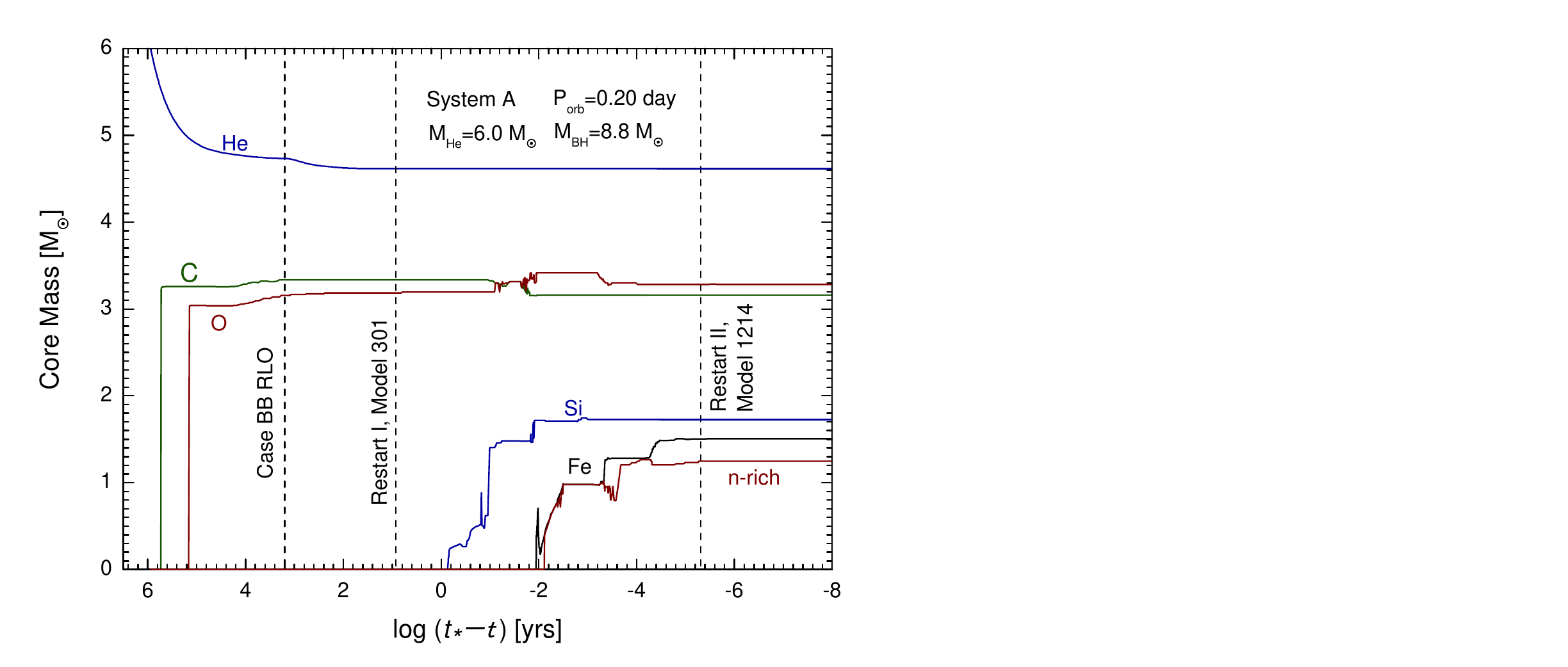}
\includegraphics[trim=1cm 7cm 3cm 0.9cm, clip=true, width=0.7\textwidth, angle=0]{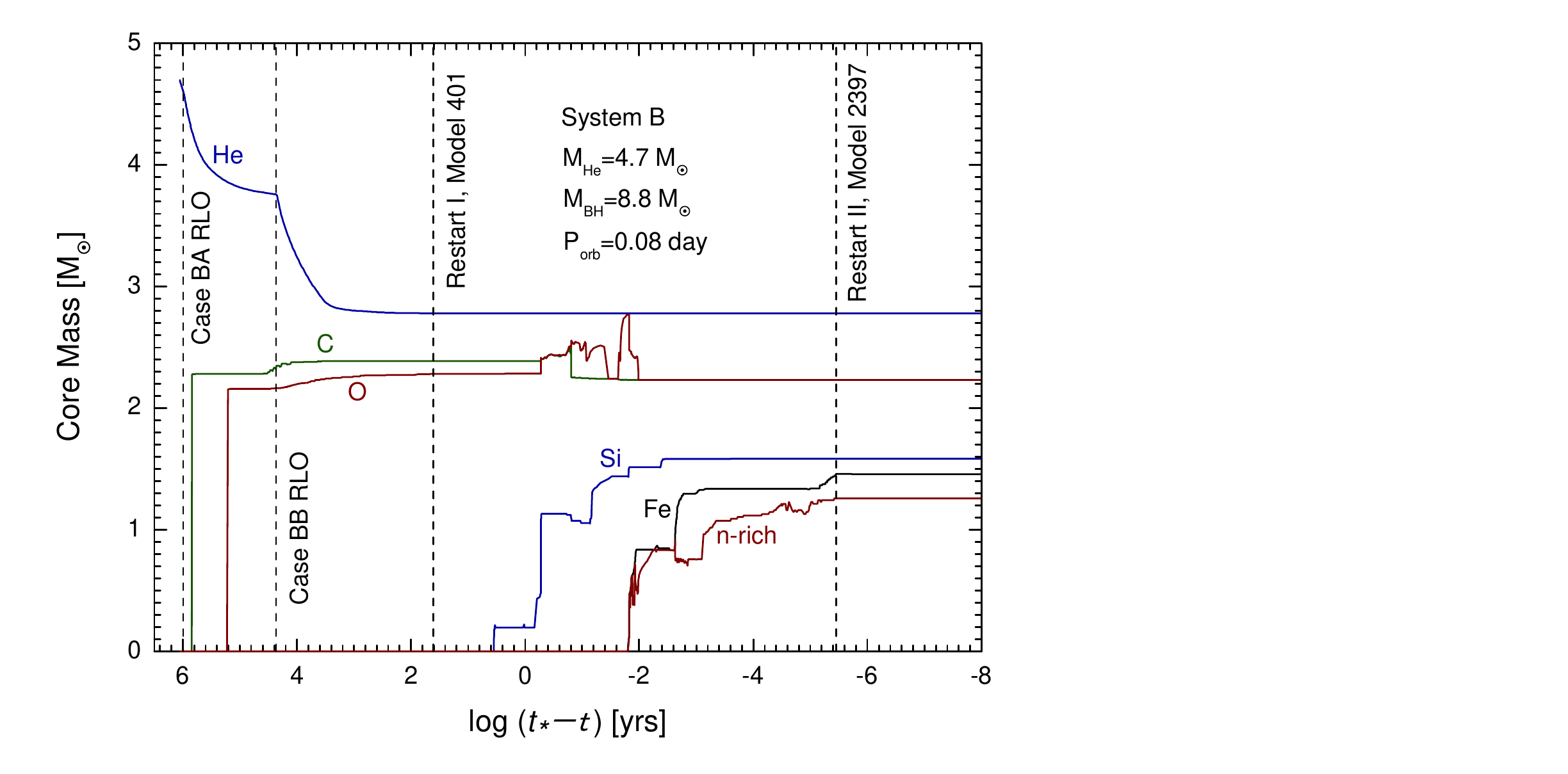}
\includegraphics[trim=1cm 7cm 3cm 0.9cm, clip=true, width=0.7\textwidth, angle=0]{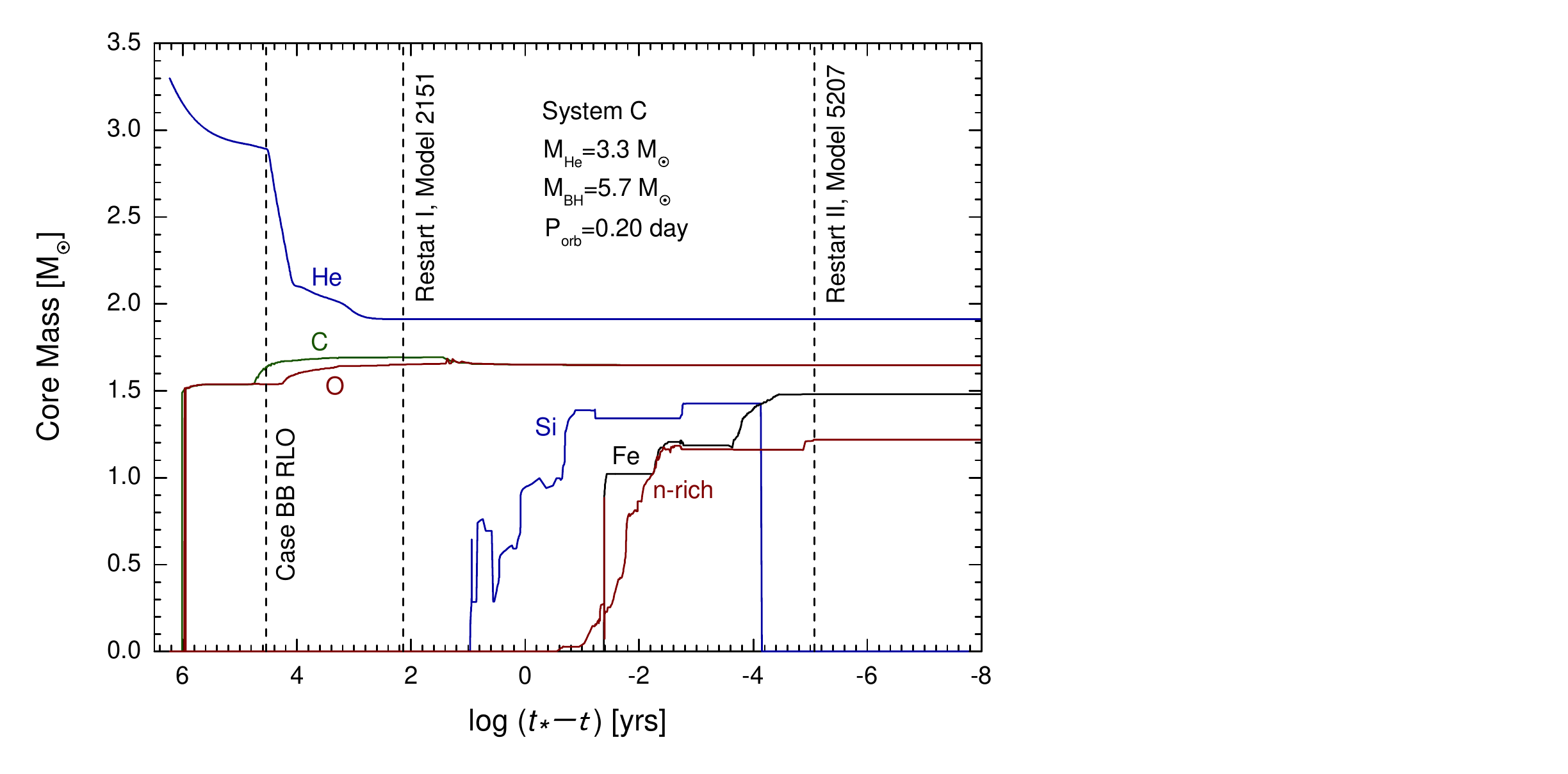}
\caption{Evolution of the mass fractions of the most abundant chemical elements of the three donor stars. 
The black dashed vertical lines represent the onset of Case~BA RLO, Case~BB RLO, Restart~I and II, respectively.} 
\label{fig:mass-core}
\end{figure}

In System~B, mass transfer is initiated with Case~BA RLO, 
while Systems~A and C, with larger initial orbital periods, initiate mass transfer with Case~BB RLO 
Because its high wind mass-loss causes the orbit to widen, the massive He~star in System~A fills its Roche lobe at a later epoch closer to the CCSN, $\log(t_\ast-t)\sim 3.2$, compared to in System~C. 
The mass-transfer rate of System~B, $|\dot{M}_{\rm tr}|$ resulting from Case~BA RLO is relatively low and comparable in magnitude to the wind mass-loss rate. 
In all three cases, once Case~BB RLO sets in, $|\dot{M}_{\rm tr}|$ increases rapidly (see also Figures~\ref{fig:kippenhann} and \ref{fig:mass-core}),
maintaining a rate of $10^{-6}-10^{-4}~M_{\odot}\,\rm yr^{-1}$.
Due to the limitation of the Eddington accretion rate, the BHs only gain a small amount of the transferred material and the amount of accreted mass in Systems~A, B, and C accumulates to about $0.001\;M_{\odot}$, $0.117\;M_{\odot}$, and $0.014\;M_{\odot}$, respectively.
It is thus seen that, as expected, in System~B the BH accretes significantly more than in Systems~A and C. This is because in addition to Case~BB RLO it also undergoes Case~BA RLO. This early phase of mass transfer lasts for $\sim 10^6\;{\rm yr}$, substantially longer than the following Case~BB RLO ($\sim 2.3\times 10^4\;{\rm yr}$). The BH in System~B accretes $\sim 0.102\;M_\odot$ from its donor during Case~BA RLO, but only $\sim 0.015\;M_\odot$ in the subsequent Case~BB RLO.

It should be noticed that, to speed up the detailed computations by reducing the number of iterations, we applied the explicit mass-transfer scheme in $\texttt{MESA}$ (rather than the implicit method) {whenever $|\dot{M}_{\rm He}|< 10^{-10}\;{\rm M}_\odot\,{\rm yr}^{-1}$}. 
The resultant values of $|\dot{M}_{\rm He}|\ll 10^{-10}\;{\rm M}_\odot\,{\rm yr}^{-1}$ that occur in Figure~\ref{fig:mass-loss-rate} within $\sim 1\;{\rm yr}$ prior to the CCSN is therefore a numerical artefact {of this method} --- in reality, we expect the donor star to detach from its Roche lobe. 
However, this does not affect the final structure of the exploding star which remains frozen in this short time interval.
In general, numerical artifacts are not uncommon in such calculations once $\log (t_\ast - t)\leq 1.0$.

\begin{figure*}
\hspace{-1cm}
\begin{tabular}{cc}
\includegraphics[trim=1cm 9cm 14cm 0.9cm, clip=true, width=0.47\textwidth, angle=0]{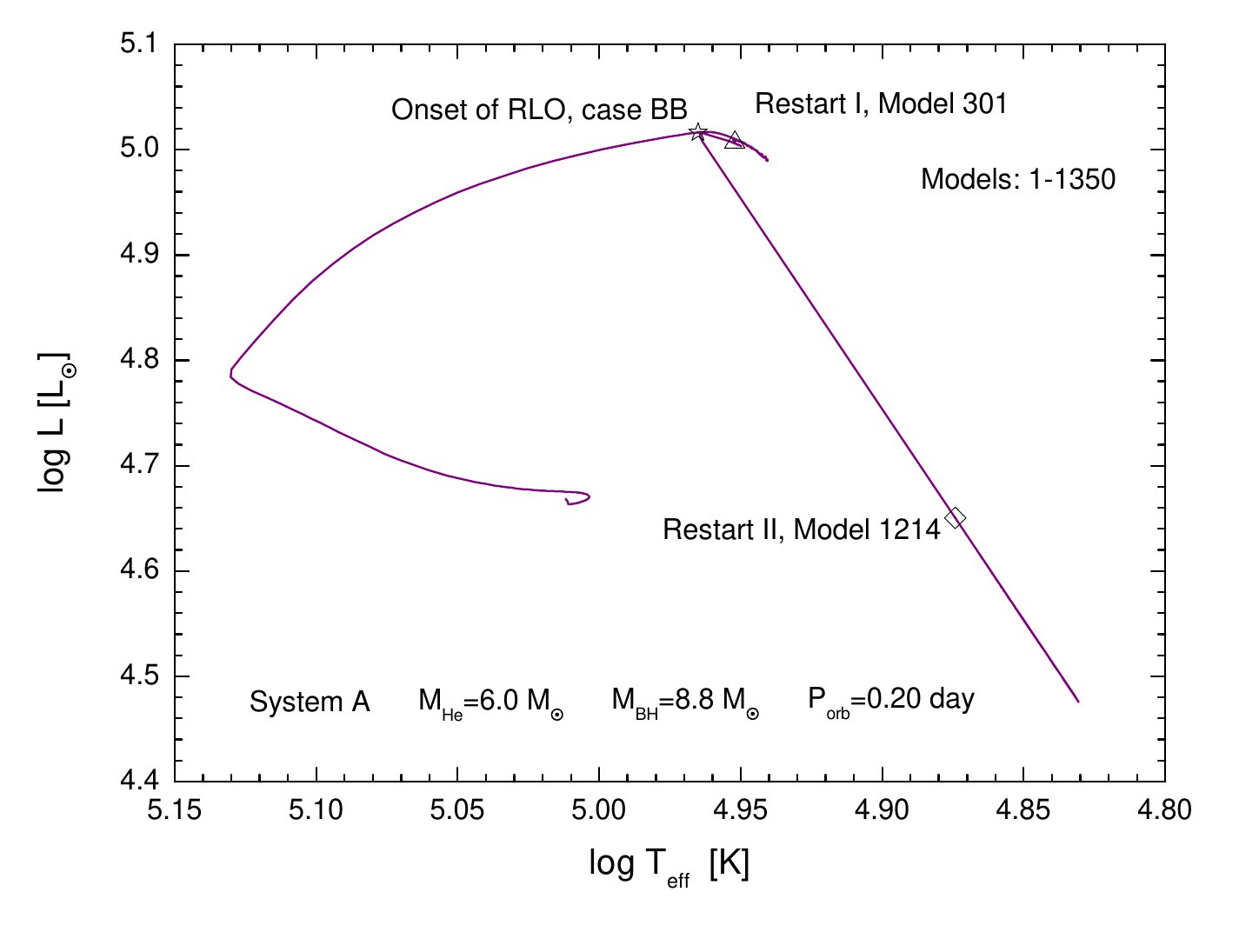} 
\includegraphics[trim=1cm 9cm 14cm 0.9cm, clip=true, width=0.47\textwidth, angle=0]{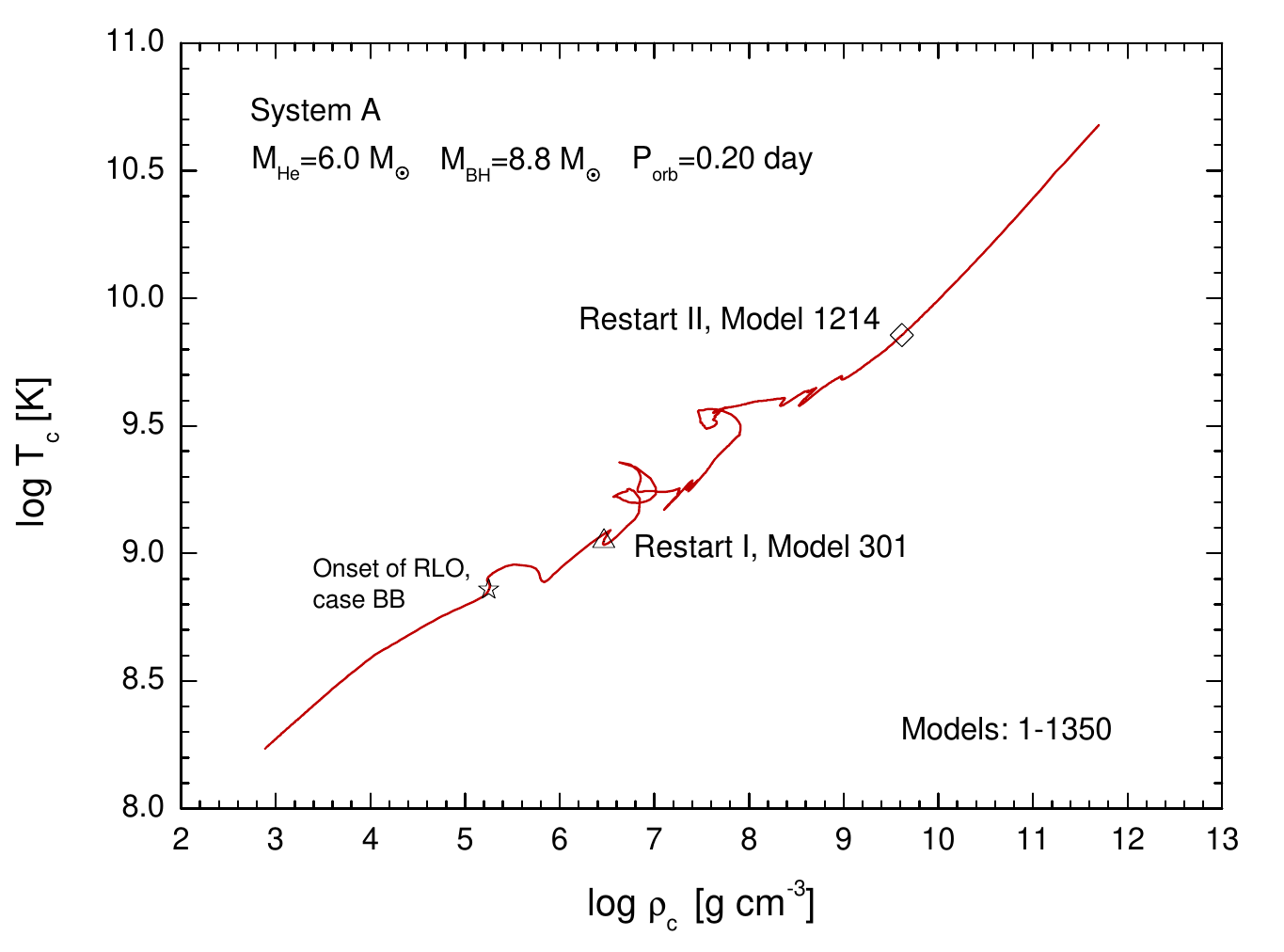}\\
\includegraphics[trim=1cm 9cm 14cm 0.9cm, clip=true, width=0.47\textwidth, angle=0]{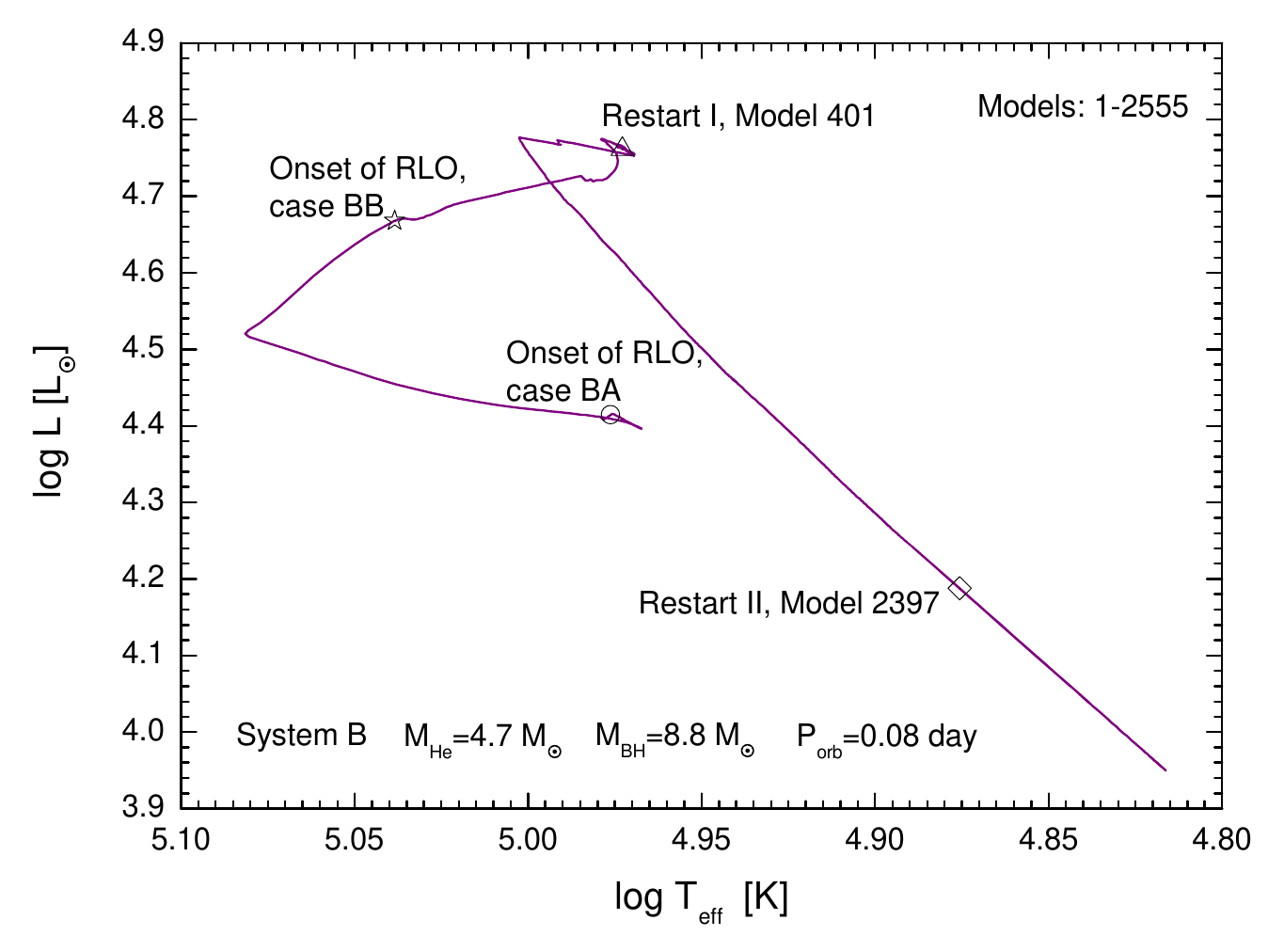} 
\includegraphics[trim=1cm 9cm 14cm 0.9cm, clip=true, width=0.47\textwidth, angle=0]{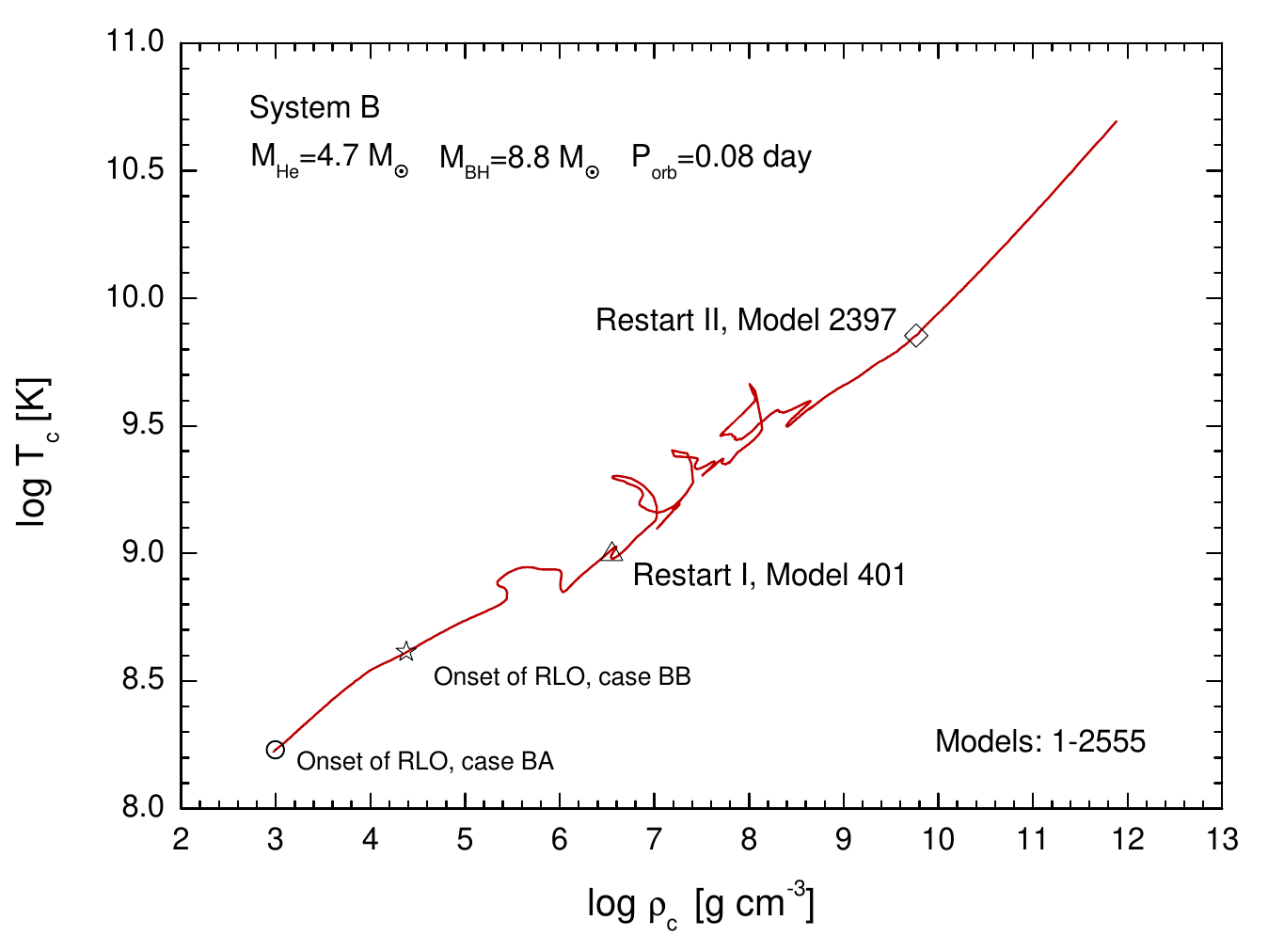}\\
\includegraphics[trim=1cm 9cm 14cm 0.9cm, clip=true, width=0.47\textwidth, angle=0]{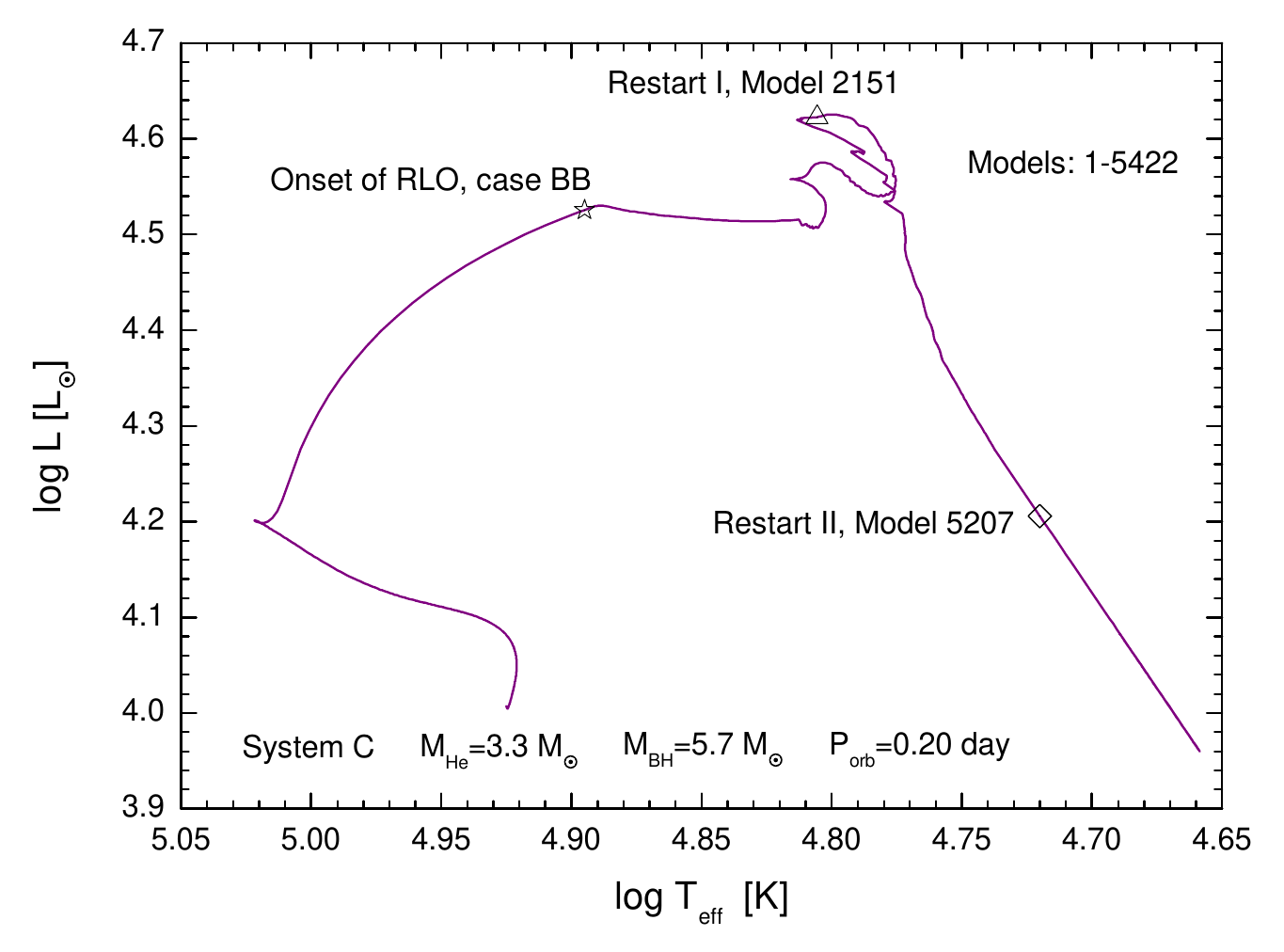} 
\includegraphics[trim=1cm 9cm 14cm 0.9cm, clip=true, width=0.47\textwidth, angle=0]{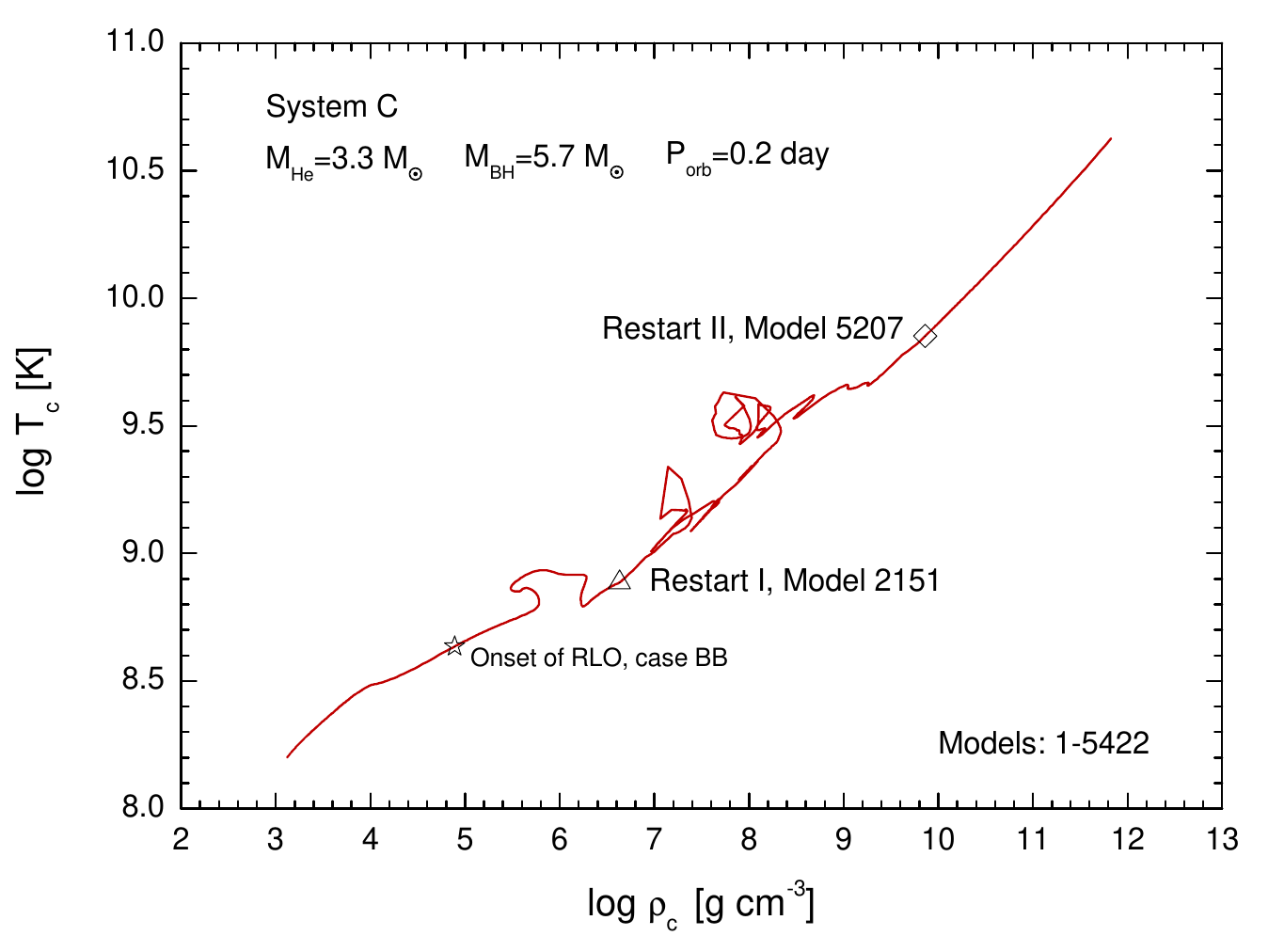}\\
\end{tabular}
\caption{Evolutionary track in the H-R diagram (left panels) and the evolution of central temperature vs. central density (right panels) of three systems. The onsets of Case~BA and Case~BB RLO, Restart~I and II are marked with circles, stars, triangles, and diamond symbols, respectively.} 
\label{fig:HR}
\end{figure*}

Figure~\ref{fig:kippenhann} shows the Kippenhahn diagram of the He stars.  
Here, the nuclear energy production rate in the core is seen to increase prior to the iron core collapse, which is also reflected in the rapidly ascending central temperature (Figure~\ref{fig:HR}, right panels). 
Figure~\ref{fig:mass-core} demonstrates the evolution of the chemical composition of the donor star cores.
The onsets of Case~BA RLO and Case~BB RLO, Restart~I and II are illustrated as black dashed vertical lines. 
The onset of Case~BB RLO (at time $\log (t_\ast - t)\simeq 4.5$) is clearly seen in Figures~\ref{fig:mass-loss-rate} and \ref{fig:mass-core} 
and marks the phase of super-Eddington mass transfer which efficiently decreases the total mass of the donor stars.

Certain ``knee'' features are seen in Figures~\ref{fig:kippenhann} and \ref{fig:mass-core} in the plotted line for the He-core mass (total mass of the donor) evolution. 
Generally, these features indicate an increase in the mass-loss rate of the donor at the onset of Case~BB RLO; and the later in evolution this begins, the less pronounced this feature is. 
Furthermore, at the point with $\log (t_\ast - t)\simeq 4.0$ in System~C, there is also a discontinuity in the evolution of the decreasing mass of the donor star (bottom panels of Figures~\ref{fig:kippenhann} and \ref{fig:mass-core}) and a sharp dip of its mass-loss rate (bottom panel of Figure~\ref{fig:mass-loss-rate}). 
These may stem from the ignition of carbon (in effect: transition from Case~BB RLO to Case~BC RLO), see the bottom panel of Figure~\ref{fig:kippenhann}.

\begin{figure}
\centering
\includegraphics[trim=1cm 9cm 3cm 0.9cm, clip=true, width=0.8\textwidth,angle=0]
{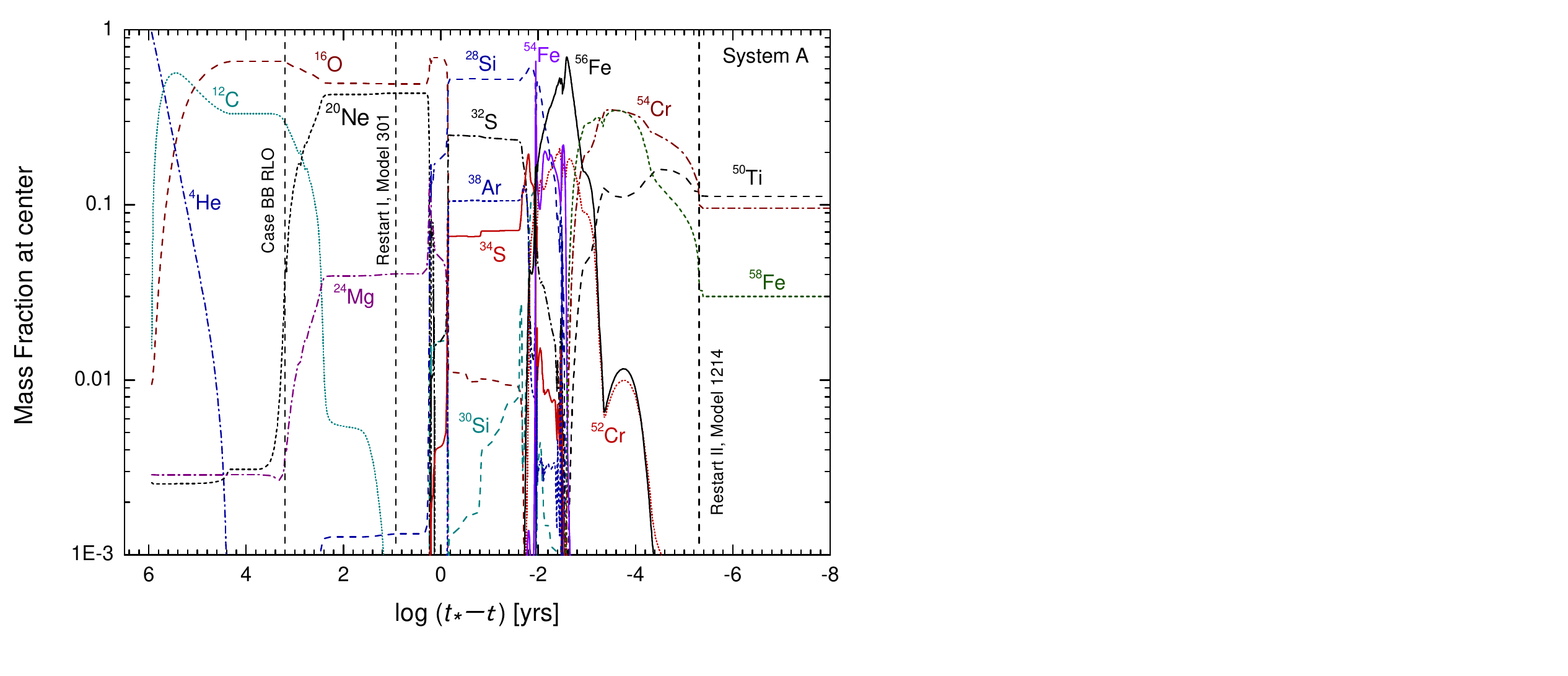}
\includegraphics[trim=1cm 9cm 3cm 0.9cm, clip=true, width=0.8\textwidth,angle=0]{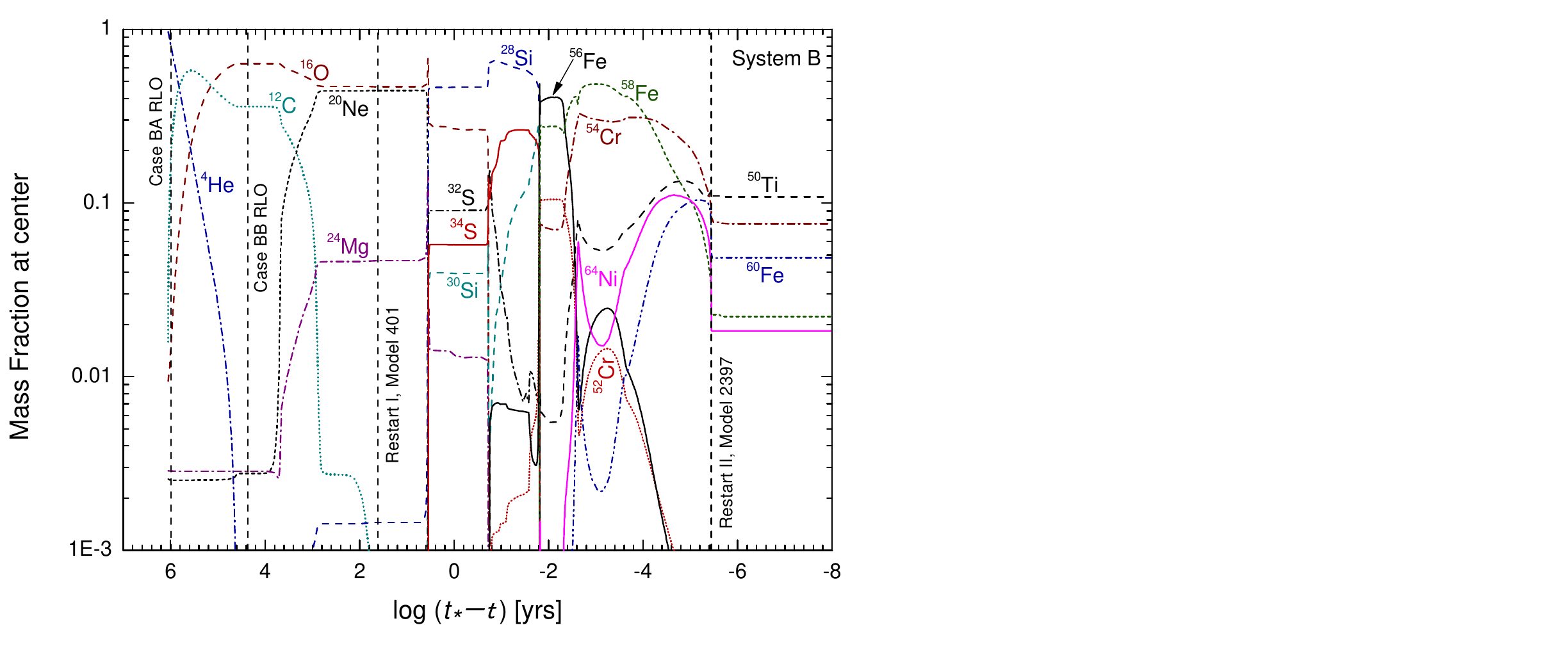}
\includegraphics[trim=1cm 9cm 3cm 0.9cm, clip=true, width=0.8\textwidth,angle=0]{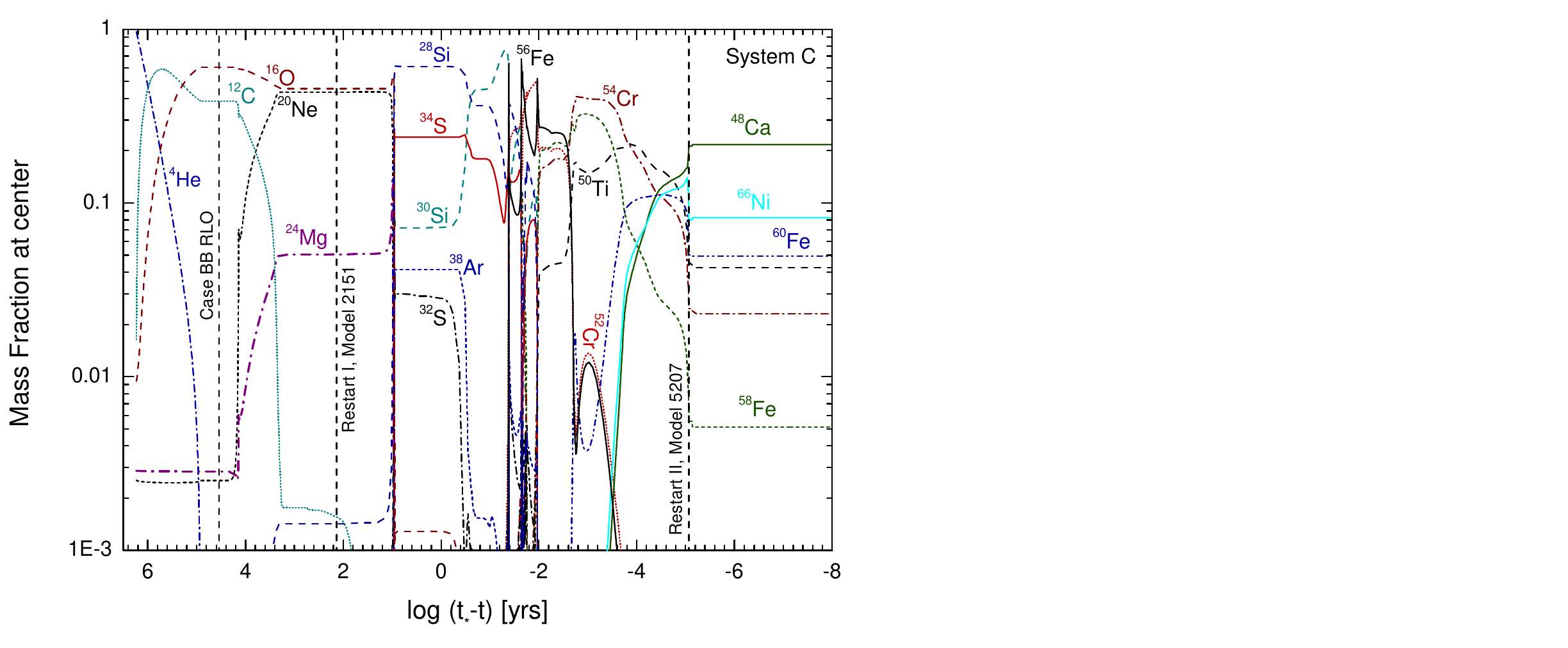}
\caption{Central mass fractions of the most abundant chemical elements vs. remaining stellar age until core collapse. The vertical lines have the same meaning as in Figure~\ref{fig:mass-core}.} 
\label{fig:element-dev}
\end{figure}

\begin{figure}
\centering
\includegraphics[trim=0.5cm 8.5cm 3cm -0.5cm, clip=true, width=0.75\textwidth, angle=0]{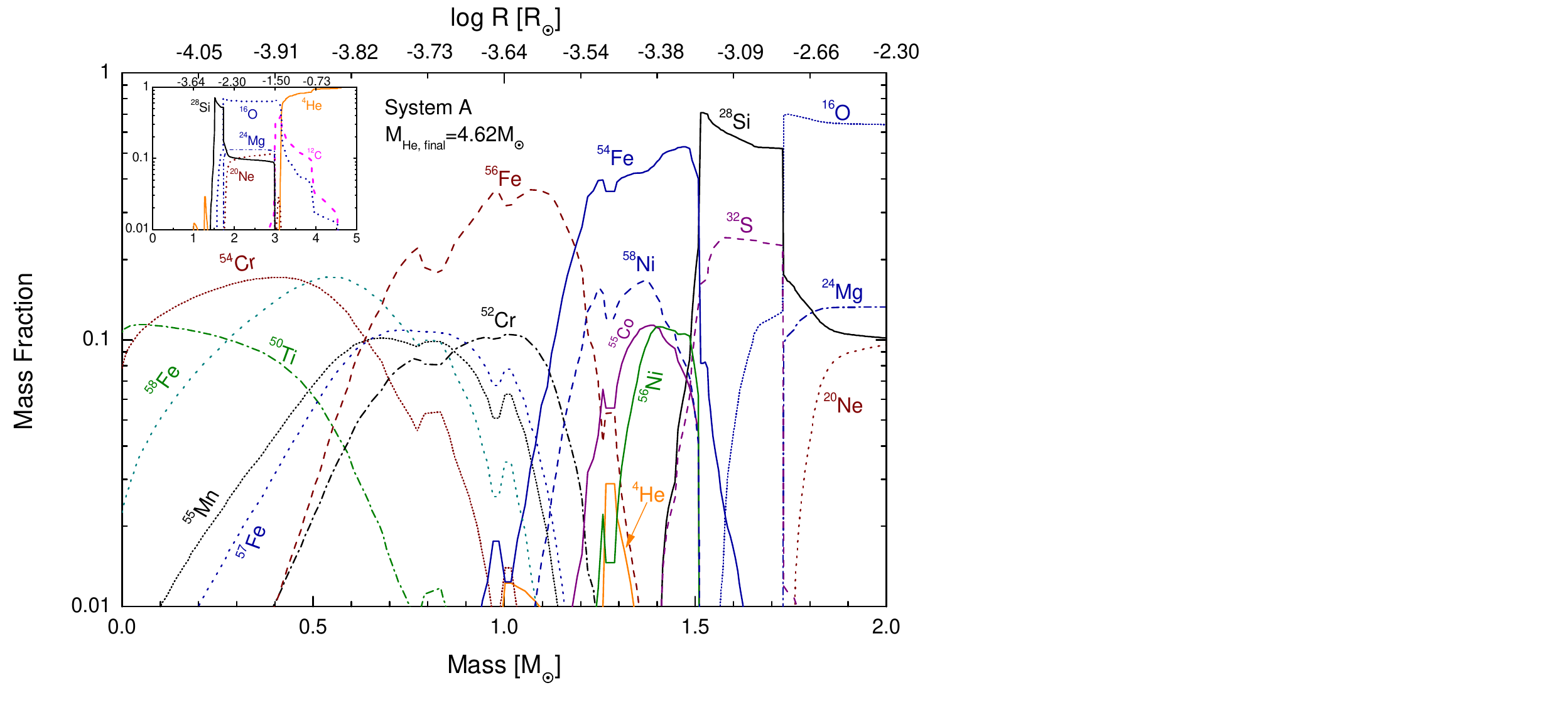}
\includegraphics[trim=1.0cm 8.5cm 3cm 0.cm, clip=true, width=0.8\textwidth, angle=0]{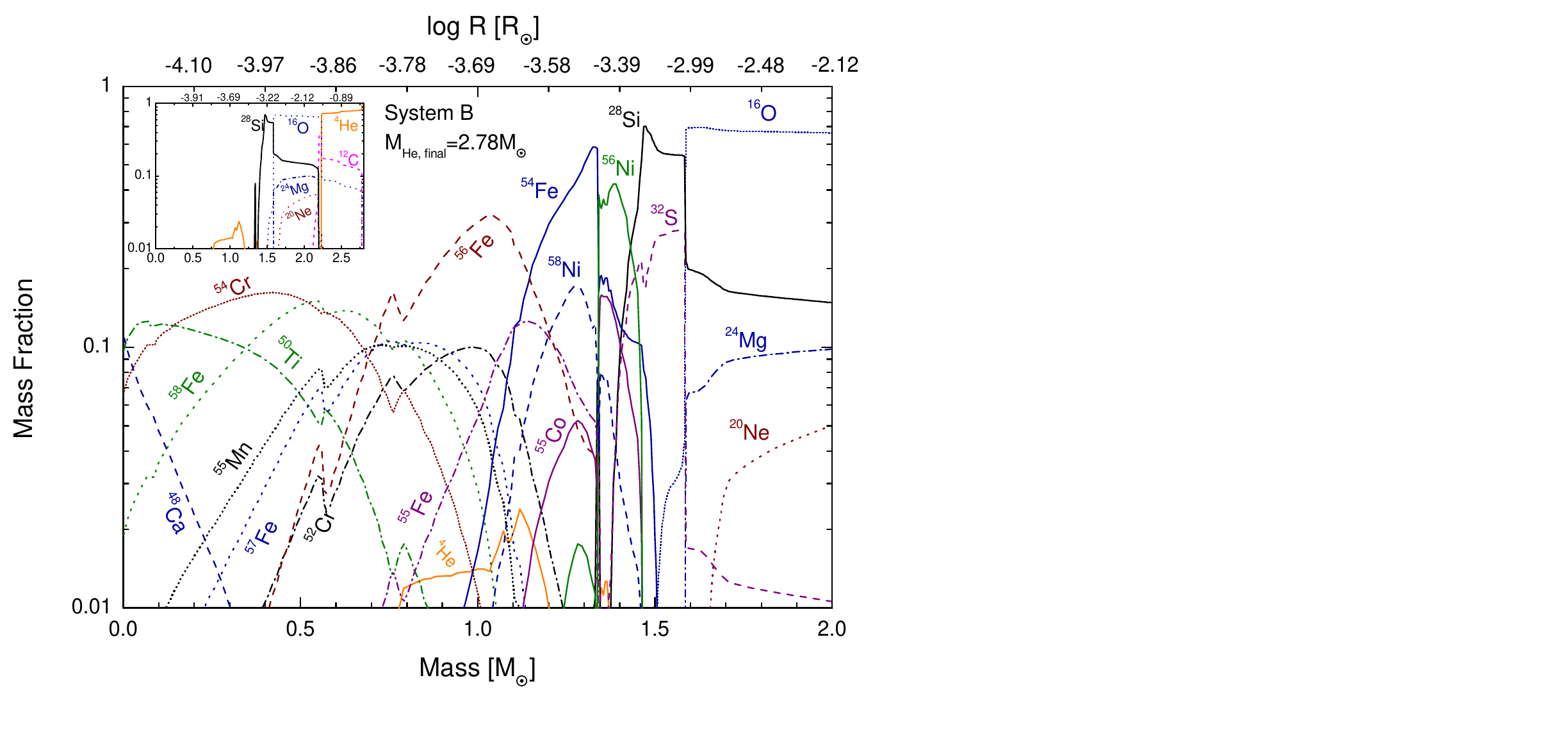}
\includegraphics[trim=1.0cm 9.0cm 3cm 0.cm, clip=true, width=0.8\textwidth, angle=0]{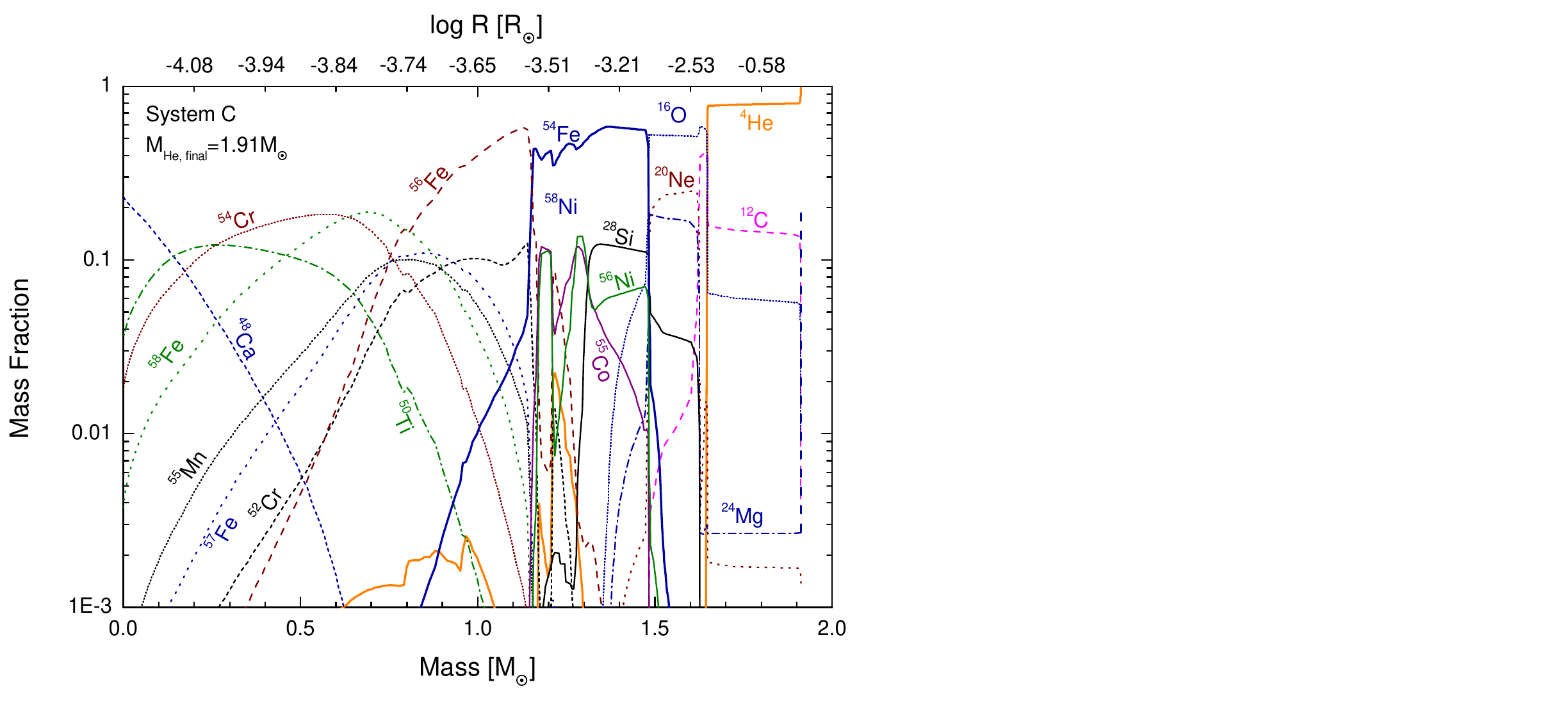}
\caption{Chemical element mass fractions as a function of mass coordinate of the (ultra-)stripped donor stars in our last simulated models at core collapse at stellar age $t_\ast$.  The radius coordinates are shown at the top. For comparison, the mass-density profiles are shown in Figure~\ref{fig:density}.} 
\label{fig:abundances}
\end{figure}

In the bottom panel of Figure~\ref{fig:mass-core}, there is a sharp decline of the silicon core mass in System~C at $\log (t_\ast-t)\sim -4.2$, 
which is caused by the core element definition in the code 
(i.e. it only defines the outer edge of the core element where its mass fraction increases to be greater than $30\%$). 
It implies that the silicon mass fraction in the core of the He~star is always 
smaller than $30\%$. 
Furthermore, at the stage of oxygen burning of small-core stars, just like System~C, electron capture leads to neutronization \citep{arne74, thie85},  which results in earlier pile-up of neutron-rich elements ($^{30}\rm Si$ and $^{34}\rm S$), and subsequently the formation of a neutron-rich core prior to the Fe-core collapse in System~C.
More details on nuclear reactions and neutron-rich elements are discussed below in connection with Figures~\ref{fig:element-dev} and \ref{fig:abundances}.

Figure~\ref{fig:HR} displays the evolutionary tracks in the H-R diagram (left panels) and the evolution of central temperature vs. central density (right panels), respectively. In each panel is marked the onset of Case~BA/BB RLO and the transition points, Restart~I and Restart~II. Upon the onset of mass transfer, the luminosities of the He~stars sharply decrease prior to iron core collapse. This is commonly seen for donor stars transferring mass at a very high rate. It is a direct consequence of the star being driven out of thermal equilibrium. In order for the star to replace the lost envelope with material from further below, and to remain in hydrostatic equilibrium, an endothermic expansion of its inner region (requiring work against gravity) causes the surface luminosity to decrease significantly. 
The rapidly ascending central temperature in the right panels in the late stages, i.e. before and after Restart~II marked with a diamond, is caused by the shrinkage of the central core (iron core infall) before CCSN.

Figure~\ref{fig:element-dev} summarizes the evolution of the most abundant elements at the center of the donors. During our simulations, a nuclear network which includes 235 nuclei is adapted. It is impossible and meaningless to trace all of their evolution.
Some elements with a very low abundance have not been presented in Figure~\ref{fig:element-dev} for clarity, which results in a fraction sum of all elements being only $\sim 0.9$.
For numerical stability purposes, we had to restart our computations twice --- see the two transition points Restart~I and Restart~II. At Restart~I, the central mass fraction of $^{28}\rm Si$ is up to $f_{\rm Si}\simeq 0.15\%$ (well before oxygen ignition), while at Restart~II the central mass fraction of $^{56}\rm Fe$ dropped to zero ($<0.01\%$). At Restart~II, the abundances of synthesized neutron-rich elements reach stable values, and the neutron-rich elements core masses have approached their maximum values (see also Figure~\ref{fig:mass-core}). 
Moreover, when the remaining stellar age is less than $\sim 1\;{\rm hr}$ ($(t_\ast-t) \la 10^{-4}\;{\rm yr}$), the combined plotted mass fractions drop to $\la 0.2$ because only 18 chemical elements are plotted.
The sharp increase of $^{48}\rm Ca$ (and the drop of $^{50}\rm Ti$) at $(t_\ast-t)\sim 10^{-5}\;{\rm yr}$ indicates the pile-up of neutron-rich elements at the center.

In the bottom panel, the feature of neutron-rich elements increasing is shown as the abrupt decrease of $^{28}\rm Si$ and increase of $^{30}\rm Si$ when the remaining time is about $\log(t_\ast-t)\simeq-0.6$.
Furthermore, the latter element (neutron-rich, $^{30}\rm Si$) shares a similar shape to the line of neutron-rich core mass in the bottom panel of Figure~\ref{fig:mass-core} at $-0.6>\log(t_\ast-t)>-1.4$. Besides, the mass fraction of $^{34}\rm S$ is higher than that of its most stable isotope $^{32}\rm S$, which is also different from the case in Systems~A and B. These can be taken as the results of electron capture at the late stage of oxygen burning \citep{arne74, thie85}.

Figure~\ref{fig:abundances} illustrates the chemical mass fractions of our final models as a function of the mass coordinate at stellar age $t_\ast$. The plotted elements were selected if their mass fractions were greater than $10\%$ at any shell. At the moment of core collapse, the innermost part of the He~star (within a mass coordinate of $\sim 1.2-1.3\;{\rm M}_\odot$) is composed of neutron-rich elements such as $^{48}\rm Ca$,  $^{54}\rm Cr$, $^{50}\rm Ti$, $^{58}\rm Fe$ and so on.
Some layers in Systems~A, B, and C are dominated by helium, oxygen, and iron. However, in System~C there are no layers dominated by silicon and nickel. This discrepancy should stem from weak interactions in the late stage of oxygen burning, which produce neutron-rich elements: $^{30}\rm Si$ and $^{34}\rm S$. Subsequently, $^{30}\rm Si$ and $^{34}\rm S$ are fused into $^{54}\rm Fe$ and $^{56}\rm Fe$ \citep[][and references therein]{woos02}, instead of $^{56}\rm Ni$. As a result, there are no layers dominated by silicon and nickel. Furthermore, in System~A no nickel layer is formed, but a relatively thick layer of $^{28}\rm Si$ and $^{32}\rm S$ is left behind which may evolve to $^{56}\rm Ni$ if element fusion continues.

\begin{figure}
\centering
\includegraphics[trim=1cm 9cm 13cm 0.9cm, clip=true, width=0.45\textwidth, angle=0]{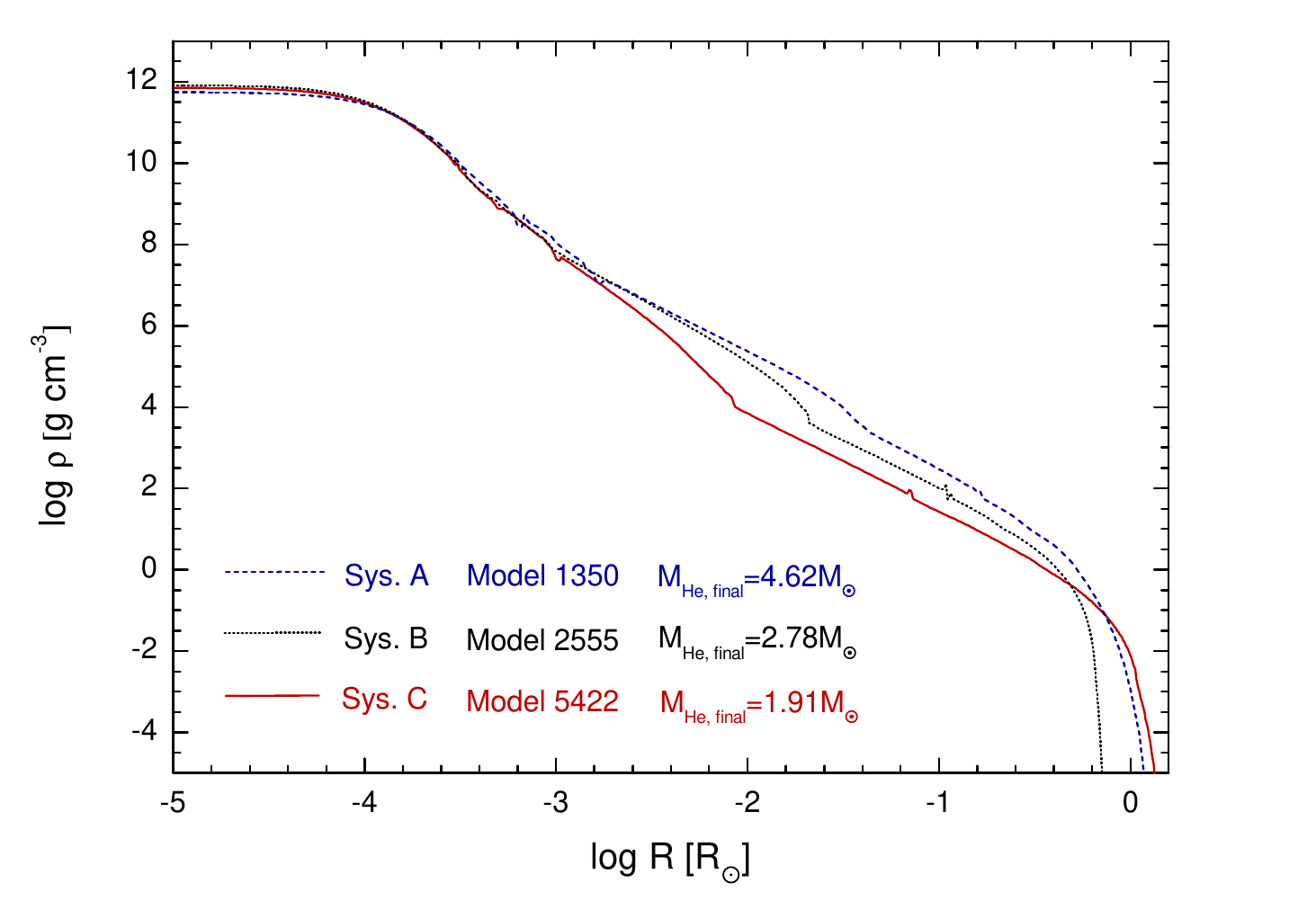}
\caption{Mass-density profiles of the three simulated (ultra-)stripped SN progenitors with stellar age $t_\ast$ at the moment of the CCSN.} 
\label{fig:density}
\end{figure}

Figure~\ref{fig:density} plots the density profile of the final models at the moment of CCSN for all three (ultra-)stripped donor stars. Inside a radius coordinate of $\sim 10^{-4}\;R_\odot$, the mass densities rise to $\sim 10^{12}{\rm ~g~cm}^{-3}$, which already exceeds the critical neutron-drip density $\rho_{\rm crit}=4.3\times10^{11}~\rm g\,cm^{-3}$ \citep{st83}. Similar to our investigation in \citet{jian21}, the simulated central densities are two orders of magnitude higher than {obtained by \citet{mori17} for a NS+NS progenitor system.}

The final masses of the SN progenitors of Systems~A, B, and C at the moment of iron core collapse with an infall velocity of $1000\;\rm km\,s^{-1}$ are {(Table~\ref{tbl-1})}: 4.55, 2.78, and $1.91\;{\rm M}_{\odot}$, respectively. These (ultra-)stripped exploding He~stars possess a CO core of 3.27, 2.23, and $1.65\;{\rm M}_{\odot}$; an iron core of 1.56, 1.46, and $1.48\;{\rm M}_{\odot}$; and a He-rich envelope of 1.28, 0.55, and $0.26\;{\rm M}_{\odot}$, respectively.

\section{Discussion}
\subsection{Final post-SN parameters}
The estimated masses of the newborn compact objects strongly depend on the exact mass cut during the SN explosion. 
To estimate the NS masses and other explosion and compact remnant properties, we first ran the SN progenitors through the semi-analytic explosion model of \citet{mueller_16}. The semi-analytic model predicts a mass cut close to the edge of the iron-silicon core in all three cases (i.e., well inside the CO core), in line with findings from many multi-dimensional simulations. The predicted gravitational NS masses are 
$1.53\;{\rm M}_\odot$ (System~A), $1.45\;{\rm M}_\odot$ (System~B), and $1.34\;{\rm M}_\odot$ (System~C).  
The explosions are expected to be of normal energy for Type~Ib/c~SNe in Systems~A and B ($1.5\times 10^{51}\;{\rm erg}$ and $1.0\times 10^{51}\;{\rm erg}$, respectively) or slightly unenergetic ($0.6 \times 10^{51}\;{\rm erg}$) for System~C. 

To estimate the final NS masses similarly to the prescription applied in \citet{jian21}, one may assume that the CO (metal) core boundary of the ultra-stripped He~star is a rough estimate for the baryonic mass cut of the newborn NS, $M_{\rm NS}^{\rm baryon}$. 
Our simulations yield CO core masses of 3.27, 2.23, and $1.65\;{\rm M}_\odot$, for Systems~A, B and C, respectively. Due to the release of gravitational binding energy, the relation between the gravitational mass and the baryonic mass of the newborn NS can be expressed as \citep{latt89}:
\begin{equation}
\label{eq:Ebind}
   M_{\rm NS}^{\rm baryon}\simeq M_{\rm NS}^{\rm grav}+0.084\;{\rm M}_\odot \;(M_{\rm NS}^{\rm grav}/{\rm M}_\odot)^2.
\end{equation}
As a result, the predicted gravitational masses of the NSs in this case for Systems~A, B, and C are thus estimated to be around 2.67, 1.92, and $1.47\;{\rm M}_\odot$, respectively. 
These values should be take as firm upper limits of the NS masses. Here it should be noted that remnant masses of $\ga 2.3\;M_\odot$ are likely to result in a BH rather than a NS \citep[e.g.][]{sfh+17}.

Given that the remnant masses predicted by the semi-analytic explosion model may be considered as lower limits, and that the results from adopting the CO core as the mass cut should provide solid upper limits, we conclude that the final mass range of the second-formed compact objects in our three families of simulated systems are $M_{\rm NS} =1.53-2.67\;M_\odot$, $1.45-1.92\;M_\odot$, and $1.34-1.47\;M_\odot$, for Systems~A, B, and C, respectively.

\subsection{Simulations with more massive helium stars}\label{subsec:massive_He-stars}
To investigate the possibility of producing more massive NS components in our final BH+NS systems, we additionally simulated systems with initial He-star masses up to $15\;{\rm M}_\odot$. As an example, a $15\;{\rm M}_\odot$ He~star accompanied by a $8.8\;{\rm M}_\odot$ BH with an initial period 0.2~days (i.e. similarly to System~A) will evolve to a NS of $1.8\;{\rm M}_\odot$, according to the mass cut of the semi-analytic model of \citet{mueller_16}. However, its final compactness parameter \citep[eq.~10 of][]{oco11} of $\xi_{2.5}\sim0.44$ is very close to the critical value of $\sim 0.45$ that produces a post-SN BH. So, this value of $1.8\;M_\odot$ could potentially indicate the upper mass limit of the newborn NS according to our simulations (although more massive NS remnants cannot be ruled out at ``islands of explodability'' further up the mass spectrum; see Section~\ref{sec:results} for references to the literature). Explosions for relatively large He or CO-core masses due to structural changes of stars in binary systems have already been noted previously by \citet{schneider_21}.

\subsection{Merger times of simulated BH+NS systems and comparison to observations}\label{subsec:merger-times}
Given that (see below) all these BH+NS systems are likely to merge and produce high-frequency GW events within a Hubble time, we notice that our example System~C is a potential progenitor of a GW200115-like event ($5.7^{+1.8}_{-2.1}\;{\rm M}_\odot,\,1.5^{+0.7}_{-0.3}\;{\rm M}_\odot$). Whereas our example systems have difficulties producing massive NSs, our subsequent simulations with more massive He~stars in System~A/B-like configurations were able to produce possible candidates of GW200105-like events ($8.9^{+1.2}_{-1.5}\;{\rm M}_\odot,\,1.9^{+0.3}_{-0.2}\;{\rm M}_\odot$).

In the rest of this subsection, we discuss the post-SN evolution of our systems.
The analytic SN model also provide an indicative scale for the NS kicks, which is 790, 465, and $260\;{\rm km\,s}^{-1}$ for example Systems~A, B, and C, respectively. The kick velocity is subject to stochastic variations in the explosion geometry, however, and may be considerably smaller if the explosion geometry is bipolar rather than dipolar \citep{mueller_19}. 

Whether or not our simulated post-SN BH+NS systems will merge within a Hubble time and produce GW events like GW200105 and GW200115 depends on the magnitude and direction of the kick imparted onto the newborn NS. 
The merger timescale of a general binary with point masses due to GW radiation is given by \citet{pet64}. 
{If, for a first rough estimate, we ignore any changes of orbital period and eccentricity due to the CCSN, one obtains for a circular post-SN binary:
\begin{equation}\label{eq:merger}
  \tau_{\rm merger}^{\rm circ}\simeq 47.1\;{\rm Gyr}\;\;\frac{(m_{\rm BH}+m_{\rm NS})^{1/3}}{m_{\rm BH}\,m_{\rm NS}}\,\left(\frac{P_{\rm orb}}{{\rm day}}\right)^{8/3},
\end{equation}
where $m_{\rm BH}\equiv M_{\rm BH}/{\rm M}_\odot$ and $m_{\rm NS}\equiv M_{\rm NS}/{\rm M}_\odot$, 
and thus Systems~A, B, and C would merge and produce GW events in timescales of 190~Myr, 82~Myr, and 2.5~Gyr (with the lower limits of NS mass), respectively. These timescales indicate a large probability that the post-SN systems would merge
within a Hubble time (especially Systems~A and B). The dynamical effect of the SN including a significant NS kick \citep[see][for a review]{tv23}, however, would in most cases increase the orbital period (thereby increasing $\tau_{\rm merger}$), but also impose an eccentricity that causes the merger time to decrease. 

Applying a range of NS kick values and calculating the resulting post-SN eccentricities based on the assumption of random (isotropic) kick directions, we performed a more detailed simulation for the merger probability of our BH+NS binaries. The calculations on the initial orbital separations and eccentricities of post-SN systems are based on eqs.~(8--11) in \citet{taur17}. Including the evolution of the eccentricity, the merger timescale is found by numerical integration using eq.~(5.14) of \citet{pet64}. Figure~\ref{fig:merger_probability} plots the probability that our resulting BH+NS binaries will survive and merge within a Hubble time (13.8~Gyr). The merger probabilities of our simulated (example) Systems~A, B, and C are 0.39, 0.72, and 0.58, respectively.

\begin{figure}
\vspace{-0.7cm}
\hspace{-0.7cm}
\includegraphics[width=0.4\textwidth, angle=-90]{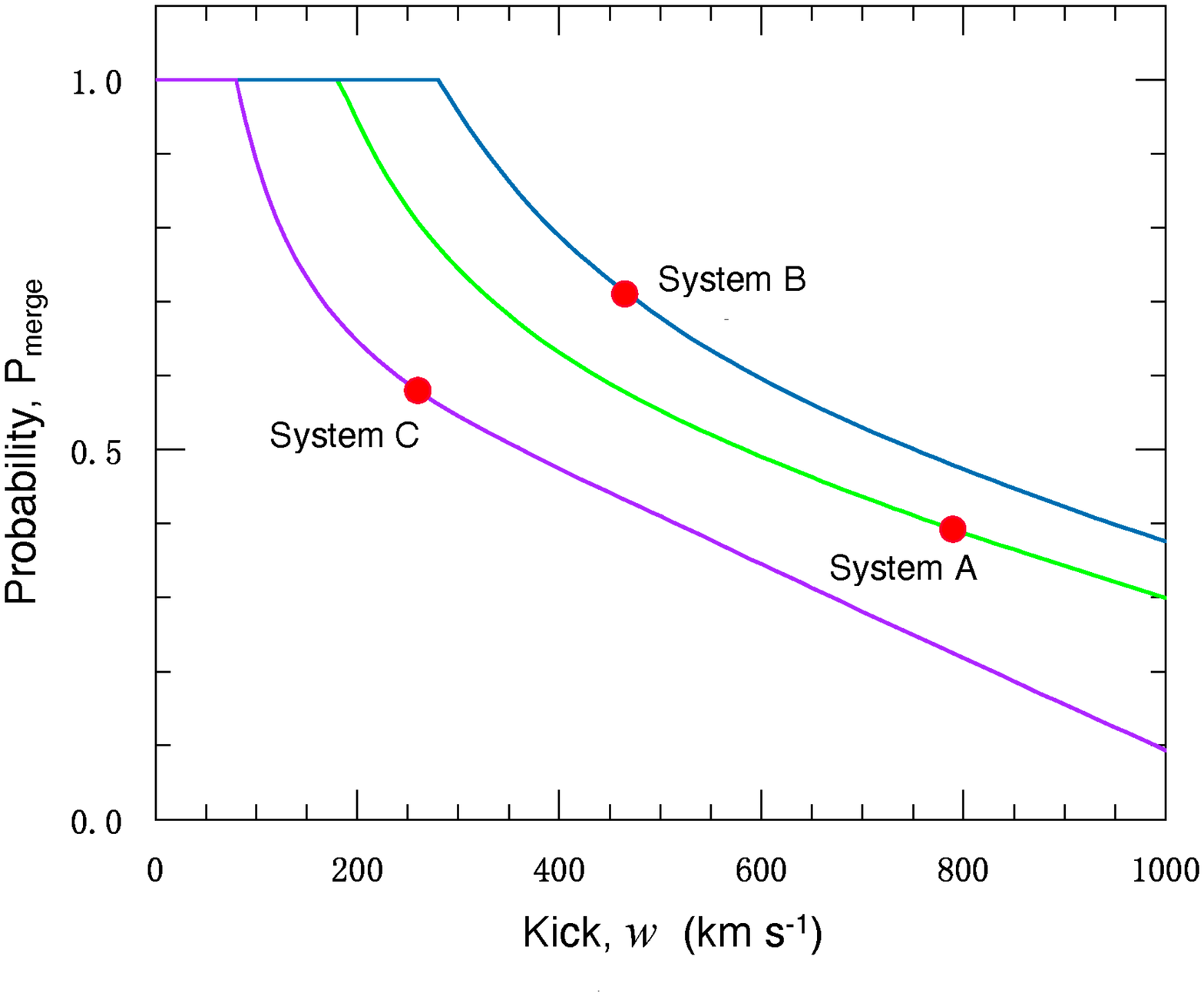}
\caption{{Probability that the post-SN BH+NS Systems~A, B, and C will merge within a Hubble time (13.8~Gyr) assuming isotropic NS kick directions as a function of NS kick magnitude.}}
\label{fig:merger_probability}
\end{figure}

In Figure~\ref{fig:merger_probability_C}, we illustrate for System~C the dependence on $\tau_ {\rm merger}$ of the kick magnitude, $w$, and the primary kick angle, $\theta$, for a fixed secondary kick angle, $\phi$ \citep[see][for a definition of the geometry]{tv23}. It is clearly seen how very different the outcome is depending on kick direction and magnitude.
\begin{figure}
\centering
\includegraphics[trim=1cm 4cm 0cm 3cm,width=0.4\textwidth, angle=-90]{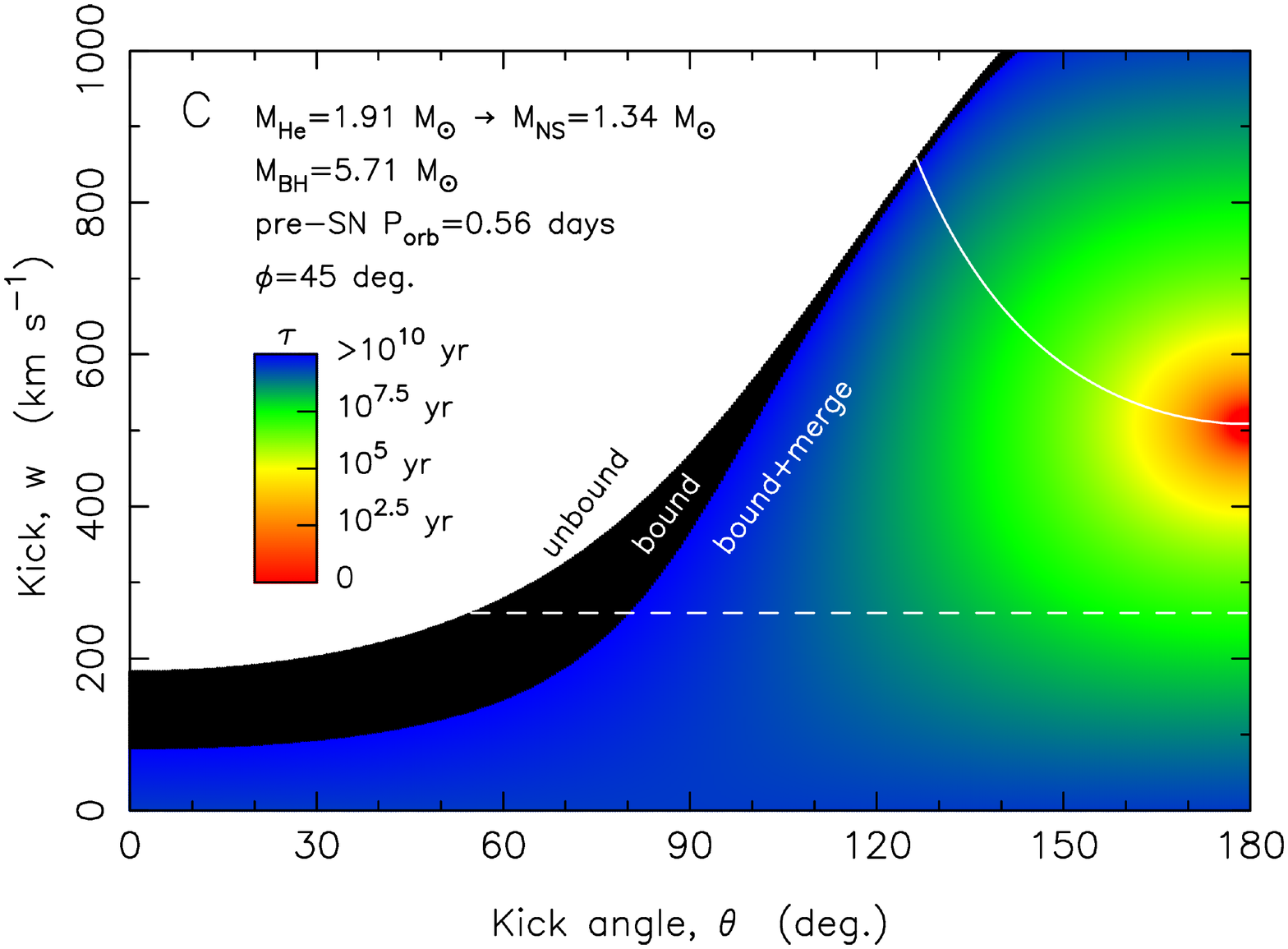}\vspace{0.0cm}
\caption{{GW merger time (color coded) of the post-SN BH+NS System~C as a function of kick angle, $\theta$ and kick magnitude, $w$. In all cases here, we assumed for illustrative purposes a constant secondary kick angle, $\phi=45^\circ$. The white area is the region where all post-SN systems disrupt because of the SN. The black-colored region right next to this area are systems that survive the SN but do not merge within 13.8~Gyr. Above the full white curve in the top-right corner, the post-SN systems become retrograde (and may result in $\chi_{\rm eff}<0$). The dashed white line marks the $w=260\;{\rm km\,s}^{-1}$ kick magnitude for System~C SN as obtained from the semi-analytic explosion model of \citep{mueller_16}.}}
\label{fig:merger_probability_C}
\end{figure}
}

\subsection{GW200115 and retrograde BH spin}
Interestingly, the BH spin of GW200115 was found to be negatively aligned with respect to the orbital angular momentum orientation \citep{abbo21}. 
Binary mergers produced via dynamical formation channels will typically possess random spin orientations \citep{mo10,rodr16}, thereby naturally resulting in a negatively aligned primary spin as observed in GW200115. However, simulations of dynamical formation channels in dense environments typically produce BH+NS merger rates much lower than that inferred from GW detections. In globular clusters and nuclear star clusters, encounter probabilities between NSs and BHs are relatively small, leading to BH+NS merger-rate densities of only $10^{-2}~\rm Gpc^{-3}\,yr^{-1}$ \citep{clau13,arca20,ye20}, which is 3 to 4 orders of magnitude lower than the empirically inferred value ($\mathcal{R}_{\rm obs}\approx 32^{+62}_{-25}~\rm Gpc^{-3}\,yr^{-1}$) and that obtained from population synthesis of isolated binary stars \citep[e.g.][]{kruc18,broe21}. 
It is also possible, however, that the progenitor binary of GW200115 was assembled well before the secondary star exploded and produced a compact object. In a young dense star cluster, tidal captures or dynamical encounters are very frequent \citep{port04,baum06}.
A massive main-sequence progenitor of the secondary He~star could be caught by a compact object via such interaction processes and subsequently become a bright X-ray source \citep{hopm04}. In case the first-formed compact object is a NS, an initial negatively aligned spin within the new binary may reverse during accretion and explain e.g. the puzzling slow spin of GR~J17480$-$2446 in the globular cluster Terzan~5 \citep{jian13}. 
If the first-formed compact object is a BH (like we investigated here), after the CE stage, the accreting BH may in principle accrete a sufficient amount of material to revert the orientation of the BH spin axis such that it aligns with the orbital angular momentum vector.
However, in our simulated systems, the BHs only accrete $\sim 0.01-0.1\;M_\odot$, which is likely insufficient to reverse their spin direction and therefore consistent with the observation of GW200115 having a negative effective in-spiral spin parameter, $\chi_{\rm eff}<0$.
Finally, it should be mentioned that the formation order of the BH and the NS remains unknown in detected GW mergers and that isolated binary star evolution may also produce retrograde spinning BHs via tossing of their spin axis during formation \citep{tau22}. 

\subsection{Connection to final fate of SS433 and GRBs}
The SS433 system consists of a Roche-lobe filling A4--7I supergiant donor star with an estimated mass of
$12.3\pm{3.3}\;{\rm M}_\odot$ and a luminosity of about $3800\;{\rm L}_\odot$, plus a compact star with a mass of $4.3\pm{0.8}\;{\rm M}_\odot$, in a 13.1~d orbital period binary \citep{hg08}. A recent measurement of the rate of orbital period increase, however, points to a BH accretor mass $\gtrsim 8\;{\rm M}_\odot$ \citep{cbdp21}. Depending on the further evolution of SS433, i.e. whether its mass transfer continues to be stable or it becomes dynamically unstable such that a CE forms, the resulting post-RLO/CE orbit of the BH+He~star binary may be sufficiently tight that it resembles our initial Systems~A and B (Table~\ref{tbl-1}). Hence, the final fate of SS433 could be BH+NS merger, depending also on the details of the second SN explosion.

The possibility to detect a gamma-ray burst (GRB) in connection with a BH+NS merger depends crucially on the spin rate of the BH, and thus on the formation order of the first-formed compact object \citep{zhu22}.
Since in our Systems~A, B, and C, the BH formed first, it is anticipated to be relatively slowly spinning and therefore the subsequent BH+NS merger event would most likely prevent tidal disruption to occur, and therefore not produce a GRB source.

\section{Conclusions}
We have performed the first complete 1D~models of BH+NS progenitor systems which are calculated self-consistently until iron core collapse. We applied the \texttt{MESA} code starting from post-CE binaries consisting of a BH and a He-ZAMS star that evolved and proceeded via mass transfer until the infall velocity of the collapsing iron core exceeded $1000\;{\rm km\,s}^{-1}$. The (ultra-)stripped SN explosion was subsequently modelled using a semi-analytic method of \citet{mueller_16} to reveal estimates of final remnant masses and momentum kicks. Using these parameters, we analysed the kinematic effects of the SN explosion.

Many population synthesis studies in the literature on isolated massive binary star evolution, with the aim to investigate the formation of tight BH+NS systems and their GW merger rates, lack the details and evidence of the final outcome that we have demonstrated in this work --- a proof-of-concept study.

Our three simulated example systems (A, B, and C) eventually evolve into BH+NS binaries with component masses of $(M_{\rm BH},M_{\rm NS}) =(8.80,\,1.53)$, (8.92,\,1.45), and $(5.71,\,1.34)\;M_\odot$, respectively, and all post-SN systems are very likely to remain bound and merge within a Hubble time. 
System~C is a potential progenitor of a GW200115-like event ([$5.7^{+1.8}_{-2.1},\,1.5^{+0.7}_{-0.3}]\;{\rm M}_\odot$), and Systems~A and B may represent the final destiny of the X-ray binary SS433.

Systems A, B and C leave BH+NS binaries with NS masses between $1.34-1.53\;M_\odot$, depending on the exact mass cut during the SN explosion. Only by applying more massive initial He~stars (e.g. $15~M_\odot$), we are able to expand the final results to BH+NS binaries with more massive NSs (e.g. $1.8~M_\odot$). We therefore conclude that such modified System~A/B-like configurations are possible progenitor candidates for a GW200105-like event.

The work presented here is a continuation of the first similar detailed study of NS+NS binaries by \citet{jian21}.  A future aim for us is a similar detailed study on producing BH+BH systems via late stellar evolution in post-CE binaries.

\acknowledgments{ We are extremely grateful to the anonymous reviewer for very constructive and detailed
comments improving our manuscript. This work was partly supported by the National Natural Science Foundation of China (under grant Nos. 11803018, 12273014, 11733009, and U2031116) and the Natural Science Foundation (under grant number ZR2021MA013) of Shandong Province, and the CAS "Light of West China" Program (grant No. 2018-XBQNXZ-B-022).}

\bibliography{ms.bbl}
\bibliographystyle{aasjournal}
\end{document}